\documentclass[epjST]{svjour}
\usepackage[sort,compress]{cite}
\usepackage{graphics}
\usepackage{graphicx}
\usepackage{amsmath,amssymb,amsfonts,bm}
\def\laco{LaCoO$_3$}
\newcommand{\mymathbf}[1]{\bm{#1}}

\newcommand{\rij}{\langle ij \rangle}

\newcommand{\bS}{\mymathbf{S}}
\newcommand{\bR}{\mymathbf{R}}

\newcommand{\bd}{\mymathbf{d}}

\newcommand{\bph}{\boldsymbol{\phi}}

\newcommand{\bk}{\mymathbf{k}}
\newcommand{\bq}{\mymathbf{q}}

\newcommand{\om}{\omega}

\newcommand{\Tr}{\text{Tr}}

\def\lsim{~\rlap{$<$}{\lower 1.0ex\hbox{$\sim$}}}
\def\gsim{~\rlap{$>$}{\lower 1.0ex\hbox{$\sim$}}}

\def\makeheadbox{{\phantom{%
\hbox to0pt{\vbox{\baselineskip=10dd\hrule\hbox
to\hsize{\vrule\kern3pt\vbox{\kern3pt
\hbox{\bfseries\@journalname\ manuscript No.}
\hbox{(will be inserted by the editor)}
\kern3pt}\hfil\kern3pt\vrule}\hrule}%
\hss}}}}

\begin{document}
\title{LDA+DMFT approach to ordering phenomena and the structural stability of correlated materials}

  \author{
    J.~Kune\v{s}\inst{1,2}\fnmsep\thanks{\email{kunes@ifp.tuwien.ac.at}} \and
    I.~Leonov\inst{3} \and
    P.~Augustinsk\'y\inst{2} \and
    V.~K\v{r}\'apek\inst{4} \and
    M.~Kollar\inst{3} \and
    D.~Vollhardt\inst{3}}
  \institute{
    Institute of Solid State Physics, TU Wien, Wiedner Hauptstr. 8, 1040 Wien, Austria \and
    Institute of Physics, Czech Academy of Sciences, Na Slovance 2, 182 21 Praha 8, Czech Republic \and
    Theoretical Physics III, Center for Electronic Correlations and Magnetism, Institute of Physics, University of Augsburg, 86135 Augsburg, Germany \and
    Central European Institute of Technology, Brno University of Technology, Technick\'a 10, 61600 Brno, Czech Republic}
  \abstract{%
     Materials with correlated electrons often respond very strongly to external or internal influences, leading to instabilities
    and states of matter with broken symmetry. This behavior can be studied theoretically either by evaluating the
    linear response characteristics, or by simulating the ordered phases of the materials under investigation.
    We developed the necessary tools within the dynamical mean-field theory (DMFT) to
    search for electronic instabilities in materials close to spin-state crossovers and to analyze the
    properties of the corresponding ordered states.
    This investigation, motivated by the physics of LaCoO$_3$, led to a discovery of
    condensation of spinful excitons in the two-orbital Hubbard model with a surprisingly rich phase diagram. The results are reviewed in the first part of the article.
    Electronic correlations can also be the driving force behind structural transformations of materials.
    To be able to investigate correlation-induced phase instabilities we developed and implemented a formalism for the computation of total energies and
    forces within a fully charge self-consistent combination of density functional theory and DMFT. Applications of this scheme to the study of structural instabilities of selected correlated electron materials such as Fe and FeSe are reviewed in the second part of the paper.
  }

\maketitle

\markboth{}{}

\section{Introduction}
\label{sec:el-inst}

The theoretical understanding of complex materials with strongly
interacting electrons is one of the most challenging areas of current research in condensed matter physics.
Experimental studies of such materials have often revealed rich phase
diagrams originating from the interplay between electronic and lattice degrees of freedom \cite{Rev1,Rev2,Rev3}. These compounds are therefore particularly interesting in view of possible technological applications.
Namely, the great sensitivity of many correlated electron materials to
changes of external parameters such as temperature, pressure, magnetic
and/or electric fields, doping, etc., can be employed to construct materials with useful functionalities.

The electronic properties of materials can be computed from first
principles by density functional theory (DFT), e.g., in the local density approximation (LDA) \cite{LDA}, the generalized gradient approximation (GGA) \cite{PB96}, or using the so-called LDA+U method \cite{Anisimov91}.
Applications of these approaches describe the phase diagrams of many simple elements and semiconductors, and of some insulators quite accurately. Moreover, they often allow to make correct qualitative predictions of the magnetic, orbital, and crystal structures of solids where the equilibrium (thermodynamic) structures are determined by simultaneous optimization of the electron and lattice systems~\cite{Baroni01,RevWIEN2k,RevVASP}.
However, these methods usually fail to describe the correct electronic
and structural properties of electronically correlated paramagnetic materials.
Hence the computation of electronic, magnetic, and structural properties
of strongly
correlated paramagnetic materials remains a great challenge.
Here the computational scheme obtained by the combination of DFT and dynamical mean-field theory (DMFT) \cite{metzner89,georges96,Kotliar04,P2:springerbook}, usually referred to as DFT+DMFT (or more explicitly as LDA+DMFT, GGA+DMFT, etc.) \cite{Anisimov97,Lichtenstein98,P2:LDADMFT2002b,Held06,Kotliar06,P2:Held07,P2:Katsnelson_RMP08,P2:FOR1346-Proceedings-2011},
provides a powerful new method for the calculation
of the electronic, magnetic, and structural properties of correlated materials
from first principles. The DFT+DMFT approach is able to give a good qualitative and quantitative
description of the properties of correlated solids in both their paramagnetic and magnetically ordered states
and demonstrates the crucial importance of electronic correlations in determining the properties of these multiband materials.
In fact, by supplementing the DFT+DMFT approach with methods of quantum information theory it even becomes possible to \emph{quantify} correlations in materials and to compare the  correlation strength of different materials, e.g., transition-metal monoxides  \cite{Byczuk12,Byczuk12_}.

The DFT+DMFT scheme makes it possible to describe and explain the effect of finite temperatures, including thermally driven phase transitions, in real materials.
By overcoming the limitations of conventional band-structure methods it opens the way for fully microscopic investigations of the structural properties of
strongly correlated systems.
So far the DFT+DMFT approach has been mainly used to calculate one-particle and local two-particle observables needed to explain experimental results obtained, for example, by
photoemission and x-ray absorption spectroscopies.
The method has been widely applied to $d$ and $f$ transition metals and their oxides or Fe-based superconducting materials.

Employing a novel implementation of DFT+DMFT in combination with plane-wave pseudopotentials it was recently demonstrated that, by performing total-energy calculations for  correlated materials, it is possible to compute atomic displacements, perform structural optimizations, determine the lattice dynamics, and explore structural transitions caused by electronic correlations~\cite{Leonov08,Leonov08_,Leonov08__,Leonov11,Leonov12,Leonov15a}. Thereby the DFT+DMFT computational scheme is able to treat electronic and structural properties of strongly correlated materials on the same footing.

By combining DFT+DMFT with linear response techniques
the
scheme can be employed to investigate both electronic and lattice instabilities of correlated materials.
In particular, it is possible to search for electronic instabilities in an unbiased manner, i.e., without prior assumptions about the broken symmetry of the ordered phase.
Lattice instabilities can be explored by evaluating a complete set of the forces acting on the atoms or the phonon dispersion relations.

In this article we review some of our results obtained during the funding period of the DFG Research Unit FOR 1346 for electronic and lattice instabilities.
The paper is organized as follows:
In Sec.~\ref{sec:electronic} we discuss the DMFT treatment of electronic (particle-hole) instabilities and states with spontaneously broken symmetry. As a pilot problem we study materials close to a spin-state crossover. This already leads to a rich phase diagram with several competing instabilities. After a general introduction to the formalism in Sec.~\ref{sec:dmft-order}, we present results for the spin-state crossover material LaCoO$_3$ in Sec.~\ref{sec:ssc}. In Sec.~\ref{sec:2bhm} we introduce a minimal two-orbital model of spin-state crossover and investigate its instabilities as well as broken-symmetry phases in Sec.~\ref{sec:exciton}.

In \ref{sec:method} we review the DFT+DMFT computational scheme for the computation of the electronic structure and phase stability of correlated materials.
In particular, we present a detailed formulation of the fully charge self-consistent DFT+DMFT scheme implemented with plane-wave pseudopotentials.
This method allows one to explore structural transformations (e.g., structural phase stability) caused by electronic correlations. We discuss  applications of the
DFT+DMFT scheme to the study of the electronic structure and phases stability of correlated materials, such as elemental iron in Sec.~\ref{sec:iron} and the parent compound of the Fe-based superconductors, FeSe, in Sec.~\ref{sec:fese}.

In Sec.~\ref{sec:forces}, we provide a detailed derivation of the DFT+DMFT method implemented within linear-response theory with respect to atomic displacements. It has been shown that this approach allows one to evaluate a full set of forces acting on the nuclei and thereby perform a structural optimization of the lattice. We  apply this new technique to study the structural phase stability of a particularly simple test material, namely solid hydrogen.
An outlook is provided in Sec.~\ref{sec:conclusions}.

\section{Electronic instabilities}
\label{sec:electronic}

  There are two general theoretical methods to explore the existence of
  long-range ordered phases with spontaneously broken symmetries:

  \emph{Method 1}: investigate a particular kind of long-range order which is assumed from the beginning, and

  \emph{Method 2}: investigate the instabilities of the normal phase.

  The first method has the obvious drawback that the broken symmetry of the ordered phase needs to be anticipated. This concerns in
  particular the translational symmetry, which is assumed in all
  calculations for extended systems. A unit cell which can host the
  ordered state is required, which may be difficult to achieve, or even
  impossible in the case of incommensurate order. Apart from the translational
  symmetry, care may also be needed to remove symmetries that might be
  present implicitly in the computational scheme or in the initial
  conditions for iterative methods such as DMFT.

\subsection{Method 1}

  Numerous forms of magnetic order of strongly correlated electron systems have
  been studied intensively by DMFT-based techniques in basic
  models~\cite{chitra99,zitzler02,sangiovanni06,hoshino10} and
  materials ranging from oxides of
  $3d$~\cite{kunes09,kunes12}, $4d$~\cite{jakobi11,kim15,dang15} or
  $5d$~\cite{arita12} elements, pnictides~\cite{yin11,aichhorn11} all the way to
  elemental metals~\cite{katsnelson99,lichtenstein01}.
  Apart from superconductivity in the attractive~\cite{keller01,capone02b,capone02,toschi05,bauer09,bauer09b,koga10}
  and multiband~\cite{koga15,vanhala15,hoshino15,hoshino16} Hubbard models
  DMFT investigations of long-range ordered states other than magnetically ordered phases are rare.
  We note that superconductivity in repulsive Hubbard model was extensively studied~\cite{maier00,maier05,haule07b}
  with cluster extensions of DMFT~\cite{dca}.

  In particular, the two-band Hubbard model proved to be simple enough to be computationally
  tractable, while exhibiting surprisingly rich physics.
  Kune\v{s} and collaborators found various ordered phases of this model
  which can be described either as spin-state ordered states or excitonic
  condensates~\cite{kunes11a,kunes14b,kunes14c,kunes16,sotnikov16}.
  An exciton is a bound state between an electron and a hole with total charge zero. 
  In this article we focus on excitons with spin $S$~$=$~$1$. Due to their bosonic 
  nature excitons can Bose condense (exciton condensation), which gives rise to 
  anomalous matrix elements of the one-particle density matrix. In the case of excitons with spin $S$~$=$~$1$
  the condensation breaks spin isotropy, while time-reversal symmetry is not necessarily 
  broken.
  Related 
  results were reported by Kaneko~{\it et al.} using the variational cluster
  approximation~\cite{kaneko14,kaneko15} and Vanhala~{\it et al.} using
  cellular DMFT~\cite{vanhala15}.
  Other examples of unconventional ordered states obtained within DMFT
  are the breaking of the channel symmetry in the two-channel Kondo
  lattice reported by Hoshino {\it et al.}~\cite{hoshino11} and the
  spin/orbital freezing scenario for
  cuprate superconductors discussed by Werner {\it et al.}~\cite{werner17a}.

  While the DMFT formalism needed for the study of broken-symmetry phases is not fundamentally different
  from that for normal (e.g., paramagnetic) phases, specific
  implementations, impurity solvers and post-processing tools need
  modifications. For example, the calculations in the exciton
  condensate reported below require the possibility of off-diagonal
  hybridization functions. For this purpose we modified the segment CT-HYB QMC
  code~\cite{werner06,alps1} to allow for a real off-diagonal
  hybridization.

  As a minor methodological development we briefly mention a trick
  adapting the maximum entropy formalism~\cite{maxent}
  for indefinite (off-diagonal) spectral densities.
  It  consists in writing the
  off-diagonal spectral function in the form
  $A(\om)=A_+(\om)-A_-(\om)$, where $A_\pm(\om)\ge 0$, which requires only minor code modifications.
  While this decomposition is highly non-unique
  we found the stability of $A(\om)$ to be satisfactory.
  An alternative approach working with an indefinite $A(\omega)$ was proposed in Ref.~\cite{otsuki17}.

\subsection{Method 2}

  The investigation of instabilities of the normal phase removes the bias towards assuming a broken
  symmetry of the ordered phase. The price to be paid is that only
  instabilities, i.e. tendencies to ordering, can be identified, while
  the detailed properties of the ordered phases and their possible
  instabilities are not accessible.
  Furthermore, when several susceptibility modes diverge simultaneously, the symmetry of the ordered
  state is not uniquely determined by the susceptibility alone. It may be determined by higher-order terms in the expansion of the free energy, see Ref.~\cite{kunes14c} for an example.
  Calculations in linear-response
  are more demanding since correlation functions need to be evaluated which are not needed for the DMFT self-consistent cycle.

\subsubsection{Computation of susceptibilities within DMFT}
\label{sec:dmft-order}

The response to a weak external field is one of the most important characteristics of any physical system. For example, it yields information about the system's stability and possible phase transitions.
For fields which couple to single-particle operators the response
of the many-body system is described by susceptibilities, i.e.,
two-particle correlation functions.
Their evaluation is a difficult and in general unsolved problem.
Computations in the framework of DMFT are simplified
by two facts: (i) susceptibilities do not enter the self-consistent solution of the DMFT equations, i.e., can be computed after a converged DMFT-solution is obtained, (ii) the essential building block, the vertex function, is local. Nevertheless, even with these simplifications the calculation of susceptibilities remains a formidable numerical task, which has been performed only in special cases so far. One example is the evaluation of equal-times correlations such as the double occupancy,
which is a standard part of DMFT implementations. However, these quantities do not contain information about the dynamics and the excitations of the system. Studies of the two-particle dynamics are usually limited to the local
(impurity) spin and charge susceptibilities, which can be determined directly from the solution
of the impurity problem \cite{mar08}. Since they describe the response to a local applied field, they do not provide information about instabilities towards long-range order.
Such information is contained in $\mymathbf{q}$-dependent susceptibilities.

  When searching for (particle-hole) instabilities within the DMFT
  formalism one is interested in static susceptibilities of the
  type
  \begin{equation}
    \label{eq:susc}
    \chi_{ij,kl}(\bq)=
    \int_0^{\beta}d\tau
    \sum_{\bR}e^{i\bq\bR}
    \bigl( \langle T
    c^{\dagger}_{\bR j}(\tau)c^{\phantom\dagger}_{\bR i}(\tau)c^{\dagger}_{\mymathbf{0} k}(0)c^{\phantom\dagger}_{\mymathbf{0} l}(0)
    \rangle
    -\langle
    c^{\dagger}_jc^{\phantom\dagger}_i
    \rangle
    \langle
    c^{\dagger}_kc^{\phantom\dagger}_l
    \rangle \big),
  \end{equation}
  where $\bq$ is a vector from the first Brillouin zone.
  In general, only the correlation functions with
  orbitals $i$ and $j$ belonging to the same atom (and similarly for
  $k$ and $l$) are relevant.  Due to the locality of the two-particle irreducible
  vertex~\cite{georges96} such functions appear on both sides of
  the Bethe-Salpeter equation and may cause a divergence. The DMFT order
  parameter is therefore local, but can exhibit a modulation in space
  (divergence at $\bq\neq0$). The susceptibility (\ref{eq:susc}) is
  obtained from the two-particle correlation function
  $\tilde{\chi}_{ij,kl}(\bq;\om_m,\om_n)$
  \begin{equation}
    \label{eq:om_sum}
    \chi_{ij,kl}(\bq)=T\sum_{m,n}\tilde{\chi}_{ij,kl}(\bq;\om_m,\om_n),
  \end{equation}
  which is the solution of  a pair of Bethe-Salpeter equations~\cite{jarrell92,georges96}
  \begin{align}
    \begin{split}
      \label{eq:bsq}
      \tilde{\chi}_{ij,kl}(\bq;\om_1,\om_2)&=\tilde{\chi}^0_{ij,kl}(\bq;\om_1,\om_2)\\
      &~~~+\,T\sum_{\om_3,\om_4}\tilde{\chi}^0_{ij,mn}(\bq;\om_1,\om_3)
      \Gamma_{mn,pq}(\om_3,\om_4)\tilde{\chi}_{pq,kl}(\bq;\om_4,\om_2)
    \end{split}\\
    \begin{split}
      \label{eq:bsi}
      \tilde{\chi}_{ij,kl}(\om_1,\om_2)&=\tilde{\chi}^0_{ij,kl}(\om_1,\om_2)\\
      &~~~+\,T\sum_{\om_3,\om_4}\tilde{\chi}^0_{ij,mn}(\om_1,\om_3)
      \Gamma_{mn,pq}(\om_3,\om_4)\tilde{\chi}_{pq,kl}(\om_4,\om_2).
    \end{split}
  \end{align}
  Here $\tilde{\chi}_{pq,kl}(\om,\om')$ is the impurity two-particle
  correlation function, and $\tilde{\chi}^0_{ij,kl}(\om,\om')$ and
  $\tilde{\chi}^0_{ij,kl}(\bq;\om,\om')$ are the local and lattice
  ``bubbles'', respectively, which are constructed from the lattice Green functions. Truncated
  summations over the Matsubara frequencies and the treatment of the
  high-frequency tails are discussed in Ref.~\cite{kunes11b}.

 DMFT calculations within linear response  were performed for one-orbital
  models in the early days of DMFT~\cite{jarrell92,jarrell95,ulmke95}.
  It is only quite recently that linear response has been used for an
  unbiased search for instabilities in two- and three-orbital Hubbard
  models~\cite{kunes14a,boehnke14,hoshino15,hoshino16,steiner16,werner17a}.
  The generalization of the above formalisms to dynamical susceptibilities
  is straightforward, but involves an analytic continuation of the
  bosonic frequency. The computation of dynamical susceptibilities within DMFT
was reported for simple models in Ref.~\cite{boehnke12} and, with a simplified RPA-like
  vertex, even for real materials~\cite{park11}.

  \subsection{Spin-state crossover: The case of \laco{}}
  \label{sec:ssc}
  The degeneracy of atomic states is a common cause for electronic instabilities in strongly correlated systems. The spin degeneracy of a ground-state atomic
  multiplet in magnetic insulators is a typical example. The vicinity of a spin-state crossover, where several atomic multiplets become
  quasi-degenerate, may give rise to complex ordering phenomena. Their physics is the subject of the following sections.

  A crossover between the low-spin state (LS) and the high-spin state (HS) (``spin-state crossover'')
  is essentially an atomic effect --- a consequence of the competition between
  the Hund's rule coupling $J$ and the crystal field (CF) splitting $\Delta$. Varying these
  parameters may lead to a level crossing~\cite{tanabe} and thus change
  the single-ion ground state.  The CF splitting is typically
  controlled by external or chemical pressure. Spin-state crossovers
  were observed in many oxides of transition metals from the
  middle of the periodic table such as MnO, Fe$_2$O$_3$, FeO, CoO,
  which were theoretically studied with the LDA+DMFT
  approach~\cite{kunes08,kunes09,shorikov10,ohta12,leonov15,leonov16}.
  In bulk materials it usually gives  rise to a smooth crossover or a
  first order transition, and often involves a sizeable change of the
  specific volume. For most materials pressures in the range of tens
  of GPa are required to induce a spin-state crossover, which
  corresponds to variations of the CF on the scale of several 100 meV.

  \laco{} is an interesting exception. In this material the parameters are fine-tuned in such a way
  that a (partial) spin-state crossover can be studied by varying the
  \emph{temperature}. In fact, the  strongly temperature
  dependent magnetic and transport properties of \laco{} and related compounds have attracted much attention and have been studied
  already for half a century~\cite{heikes64,naiman65,raccah67}. Below 50 K \laco{} appears to be a  band
  insulator. However, above 100 K it exhibits a magnetic
  response typical for local moments while the charge gap continuously disappears
  between 450 and 600 K. This suggests that the material is much more
  than an ordinary band insulator.

Traditionally several different approaches have been employed to explain the
  physics of \laco{}: (i) the single-ion picture of a LS, $S$~$=$~$0$,
  ground state of the Co$^{3+}$ ion, with intermediate spin (IS), $S$~$=$~$1$,
   or HS, $S$~$=$~$2$, excitations augmented by spin-exchange
  between these states on the lattice, (ii) band structure approaches
  with electrons interacting via a static mean-field, and (iii) a
  combination of both in terms of DMFT.  The
  technically simplest approach, (i), has been used extensively to
  interpret the experimental $T$-dependence of the magnetic
  susceptibility and specific heat. It describes the physics of a
  thermally induced statistical mixture of different atomic
  multiplets, but does not capture the extended nature of a bulk
  material.  Models including IS, HS or both can be found in the
  literature~\cite{yamaguchi96,zobel02,kyomen03,kyomen05}. Generally, it is not possible to describe \laco{} by a
  single-ion model with constant parameters.
  Band structure methods, (ii), have been used to describe the
  one-particle spectra in the low-$T$ regime~\cite{abbate93,korotin96,knizek05}.  A Hartree-Fock-like LDA+U
  method has been employed to study stable spin states and their
  possible ordered patterns~\cite{hsu09,knizek09}.  However, these methods cannot capture the
  temporal fluctuations between atomic multiplets and are
  by construction limited to electrons at $T=0$. The physical
  temperature enters the calculations only via the thermal expansion
  of the lattice.
  The LDA+DMFT approach, (iii), includes the best features of the approaches (i) and (ii), and
 combines the extended nature of the systems with local
  electronic correlations~\cite{eder10,krapek12,zhang12,karolak15}.

\begin{figure}
    \includegraphics[width=0.8\columnwidth,clip]{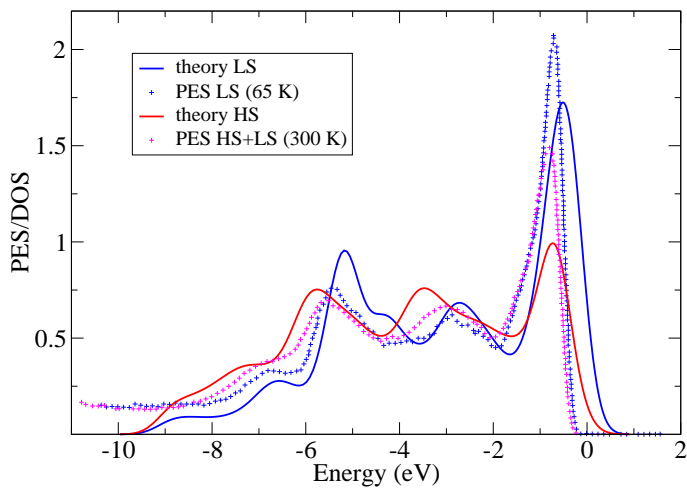}
    \caption{\label{fig:laco_exp}Comparison of the calculated spectral
      functions (lines) of \laco{} with  measurements by photoemission spectroscopy (PES)
      (symbols)~\cite{koethe06}.  The LS solution corresponds to
      $T=580$~K and the HS solution was
      obtained for $T=1160$~K. The
      measurements were taken at 65 K (denoted as LS) and 300 K
      (denoted as HS+LS, since the temperature is not high enough for the
      full spin-state crossover). Adopted from Ref.~\cite{krapek12} with
      the permission of the authors.}
  \end{figure}

  K\v{r}\'apek {\it et al.}~\cite{krapek12} used LDA+DMFT to study the
  temperature dependence of the atomic spin state and the one-particle
  spectra. In Fig.~\ref{fig:laco_exp} their results are compared
  with the experimental photoemission spectra.
  As expected the results are quite sensitive to the choice
  of the parameters used in the computation, in particular the Hund's exchange
  $J$. The thermal expansion of the lattice has a non-negligible effect
  on the electronic properties, but by itself it cannot explain the
  experimental susceptibility. On the other hand, even with a rigid lattice one could observe the effect of thermal population of
  the spinful atomic states. Perhaps the most important result is the
  negligible contribution of the IS state, irrespective of the values of
  the interaction and double-counting parameters. This result is
  consistent with single atom calculations~\cite{tanabe,haverkort06},
  which yield the LS-HS or HS-LS sequence of the multiplet states in
  the vicinity of the spin-state crossover, but never a LS-IS sequence
  with a small gap.

\begin{figure}
    \includegraphics[width=0.8\columnwidth,clip]{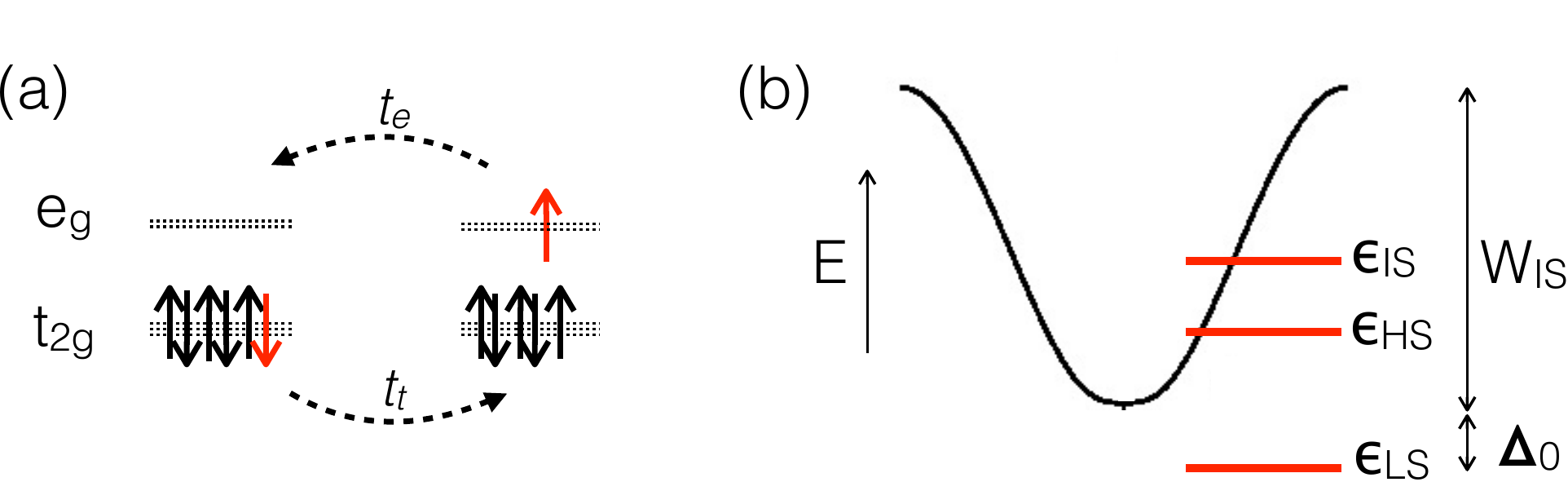}
    \caption{(a) Nearest-neighbor hopping process giving rise to
      IS-LS exchange. (b) Schematic plot of the atomic multiplet energies
      together with the dispersion of a single IS state on the LS
      background. After Ref.~\cite{sotnikov16}}
    \label{fig:cartoon1}       
  \end{figure}

  Recently, Sotnikov and Kune\v{s}~\cite{sotnikov16} proposed a
  mechanism in which the IS states still play an important role in
  \laco{}.  Starting from a global LS ground state, the IS states can be
  viewed as tightly bound excitons which carry spin $S$~$=$~$1$ and appear in
  three orbital flavors. A perturbation expansion in the
  hopping~\cite{sotnikov16} shows that the excitons have a rather high
  anisotropic mobility, i.e.,  each orbital flavor moves predominantly in
  one of the three cubic planes.  The HS states are then viewed as tightly
  bound bi-excitons with total spin $S$~$=$~$2$ formed by two IS states
  with different orbital flavors.  The estimated mobility of HS
  bi-excitons is much smaller than that of IS excitons.  The
  proposed excitation spectrum of \laco{} at low temperature is shown in
  Fig.~\ref{fig:cartoon1}.
  The energies of atomic multiplets follow
  the LS-HS-IS sequence.
  However, the dispersion of IS excitons causes the lowest excitation on the lattice to be an
  IS wave with a specific crystal momentum $\bq$.

  This picture is clearly beyond the capabilities of the single-ion model,
  (i), as well as band structure approaches, (ii).  The question is then whether DMFT is able to capture the
  proposed behavior. Although the answer turns out to be negative, it can provide useful
  insights. Roughly speaking DMFT sees the IS excitation at its atomic
  $\bq$-averaged energy, i.e., substantially above the HS state.  It
  does not take into account that in parts of the Brillouin zone
  propagating excitons have substantially lower energy.  This
  explains the results of K\v{r}\'apek {\it et al.}~\cite{krapek12}.
  A linear response calculation within DMFT performed on top of the
  low-temperature LS solution should be able to identify the
  dispersive excitonic branch --- at least its low-energy part that can
  be distinguished from the particle-hole continuum. Such a
  calculation, while possible in principle, will be quite demanding and has yet
  to be done. Altogether one may say that DMFT can describe the dispersion
  of a single IS excitation on top a LS state, but cannot describe the
  state in which such (strongly interacting) excitations are thermally
  populated.

  Experimentally, an observation of the low-energy excitonic branch in
  the low-temperature (LS) state of \laco{} based on two-particle
  spectroscopy, e.g., resonant x-ray scattering, offers a way to test
  the proposed scenario. It is interesting to note that on the
  one-particle level stoichiometric \laco{} at low temperature
  appears to be an uncorrelated band insulator with sharp
  quasi-particle bands~\cite{krapek12}.
  The ``hidden'' correlations in the material become manifest in the
  one-particle spectra only upon heating or doping.

  \subsection{Two-orbital Hubbard model}
  \label{sec:2bhm}
 In the investigation of electronic instabilities in the vicinity of a spin-state crossover, such as in cobaltites, there are two reasons not to start with realistic
 models that include the full $d$ shell. The first one is purely technical. Linear-response calculations are very demanding and their feasibility for
 the full $d$ shell with the present (still developing) codes is questionable. The second reason has to do with the interpretation of the results. As shown by
 recent LDA+U calculations~\cite{afonso16}, the excitonic order parameter in $d^6$ systems has 18 components, and numerous almost
 degenerate phases are possible. We find it therefore necessary to start with simpler systems and gradually increase the complexity.

 The two-orbital (i.e., two-band) Hubbard model (2BHM) provides such a minimal microscopic model to study
  the competition between CF splitting and Hund's rule exchange without the orbital degeneracy of realistic $d$ shell. The
  Hamiltonian reads
  \begin{equation}
    \begin{aligned}
      \label{eq:2bhm}
      H&=\frac{\Delta}{2}\sum_{i,\sigma} \left(n^a_{i\sigma}-n^b_{i\sigma}\right)+
      \sum_{\rij,\sigma}
      \begin{pmatrix}a_{i\sigma}^{\dagger},b_{i\sigma}^{\dagger}\end{pmatrix}
      \begin{pmatrix} t_a & V_{ab} \\ V_{ba} & t_b \end{pmatrix}
      \begin{pmatrix} a^{\phantom\dagger}_{j\sigma} \\ b^{\phantom\dagger}_{j\sigma} \end{pmatrix}
      +\text{h.c.}\\
      &+\,U\sum_{i} \left(n^a_{i\uparrow}n^a_{i\downarrow}+n^b_{i\uparrow}n^b_{i\downarrow}\right)+
      U'\sum_{i,\sigma\sigma'} n^a_{i\sigma}n^b_{i\sigma'}
      -J\sum_{i\sigma} \left(n^a_{i\sigma}n^b_{i\sigma}
        +a_{i\sigma}^{\dagger}a_{i\bar{\sigma}}^{\phantom\dagger}b_{i\bar{\sigma}}^{\dagger}b_{i\sigma}^{\phantom\dagger}\right)\\
      &+\,J'\sum_{i} \left(a_{i\uparrow}^{\dagger}a_{i\downarrow}^{\dagger}b_{i\downarrow}^{\phantom\dagger}
        b_{i\uparrow}^{\phantom\dagger}+\text{h.c.}\right),
    \end{aligned}
  \end{equation}
  with the notation $\bar{\sigma}=-\sigma$. It describes electrons with 
  two orbital ($a$ and $b$) and two spin
  ($\sigma=\uparrow,\downarrow$) flavors moving on a lattice and
  interacting via an on-site interaction $H_{\text{int}}$. Here,
  $a_{i\sigma}^{\dagger}$ ($a^{\phantom\dagger}_{i\sigma}$) is an
  operator creating (annihilating) a fermion with orbital flavor $a$ and
  spin $\sigma$ on the lattice site $i$,
  $n_{i\sigma}^a=a_{i\sigma}^{\dagger}a^{\phantom\dagger}_{i\sigma}$
  is the corresponding local density operator, and analogously for the
  $b$ fermions. The sum $\sum_{\rij}$ runs over the nearest neighbour
  (nn) bonds. Here we consider not only the hopping between orbitals with
  the same flavour (hopping amplitude $t$), but explicitly include also the cross-hopping between
  orbitals with different flavors (hopping amplitude $V$).
  The parameter $\Delta$ describes the CF splitting.
  The standard Slater-Kanamori interactions are
  parametrized by $U$ and $J$ (with $U'=U-2J$, $J'=J$). Numerical quantum
  Monte-Carlo calculations are greatly simplified when the interaction is
  approximated by a density-density interaction, whereby spin-flips
  (the second contribution in the $J$ term) and pair-hopping $J'$ are neglected. This approximation will be used here.

  In the following we will discuss the 2BHM at, and close to, half-filling.
  Let us first summarize the basic properties of its
  normal phase.  The physics at strong and intermediate coupling is
  controlled by the competition between the Hund's coupling $J$ and the
  crystal-field splitting $\Delta$~\cite{werner07,suzuki09}. Large
  $\Delta$ favors the singlet LS state, while large $J$
  favors the triplet HS state. When spontaneous symmetry breaking
  is excluded the $\Delta-U$ phase diagram at fixed $J/U$ can be
  divided into three regions: HS Mott insulator, LS band insulator, and
  a metallic state (``metal'')~\cite{werner07}.
  The first-order metal-insulator transition turns into a
  crossover at higher temperatures.  The low-energy physics deep in
  the Mott phase is described by the $S$~$=$~$1$ Heisenberg model with
  antiferromagnetic interaction.  The LS band insulator far from the
  phase boundaries is a global singlet with a gapped excitation
  spectrum. In the vicinity of the HS-LS crossover both LS and HS
  states have to be taken into account. The near degeneracy of the
  atomic multiplets in this region gives rise to several
  instabilities.

  Kune\v{s} and K\v{r}\'apek~\cite{kunes11a} reported spin-state order (a checker-board
  pattern of HS and LS states) on a square lattice in the 2BHM with very
  asymmetric (narrow/wide) bands.
  An interesting
  feature of this transition, which can be well explained by the
  classical Blume-Emery-Griffiths model~\cite{BEG},
  is its reentrant character in parts of the
  phase diagram~\cite{hoston91,kunes15}.
  The above (classical) order is not the only instability of the
  model. Magnetic ordering of the HS states and
  condensation of spin-triplet excitons~\cite{halperin68b,zocher11,brydon09a,brydon09b,kaneko13,kaneko14,kunes14a,hoshino16} are competing instabilities.
  In the following, we focus on the latter one.

  \subsection{Spin-triplet exciton condensation}
  \label{sec:exciton}
Long-range order in materials with singlet atomic ground states is rare.
If the energy of an atomic excitation is comparable to
its amplitude to propagate to neighbouring atoms, spontaneous symmetry breaking may take place.
This mechanism, called exciton condensation or excitonic magnetism,
was recently proposed to be realized in $4d^4$ materials such as
Ca$_2$RuO$_4$~\cite{khaliullin13,jain15} or $3d^6$ cubic materials such as the
Pr$_{0.5}$Ca$_{0.5}$CoO$_3$ family~\cite{kunes14b}.
Excitonic magnetism has been studied in simple lattice
models such a 2BHM or effective pseudospin models~\cite{khaliullin13,kunes14a,kunes14b,kaneko14,kaneko15,chaloupka16}.
In the following we review the DMFT results for the spin-triplet exciton condensate.

  \subsubsection{Linear response}
\begin{figure}
    \begin{center}
      \includegraphics[width=0.17\columnwidth,clip]{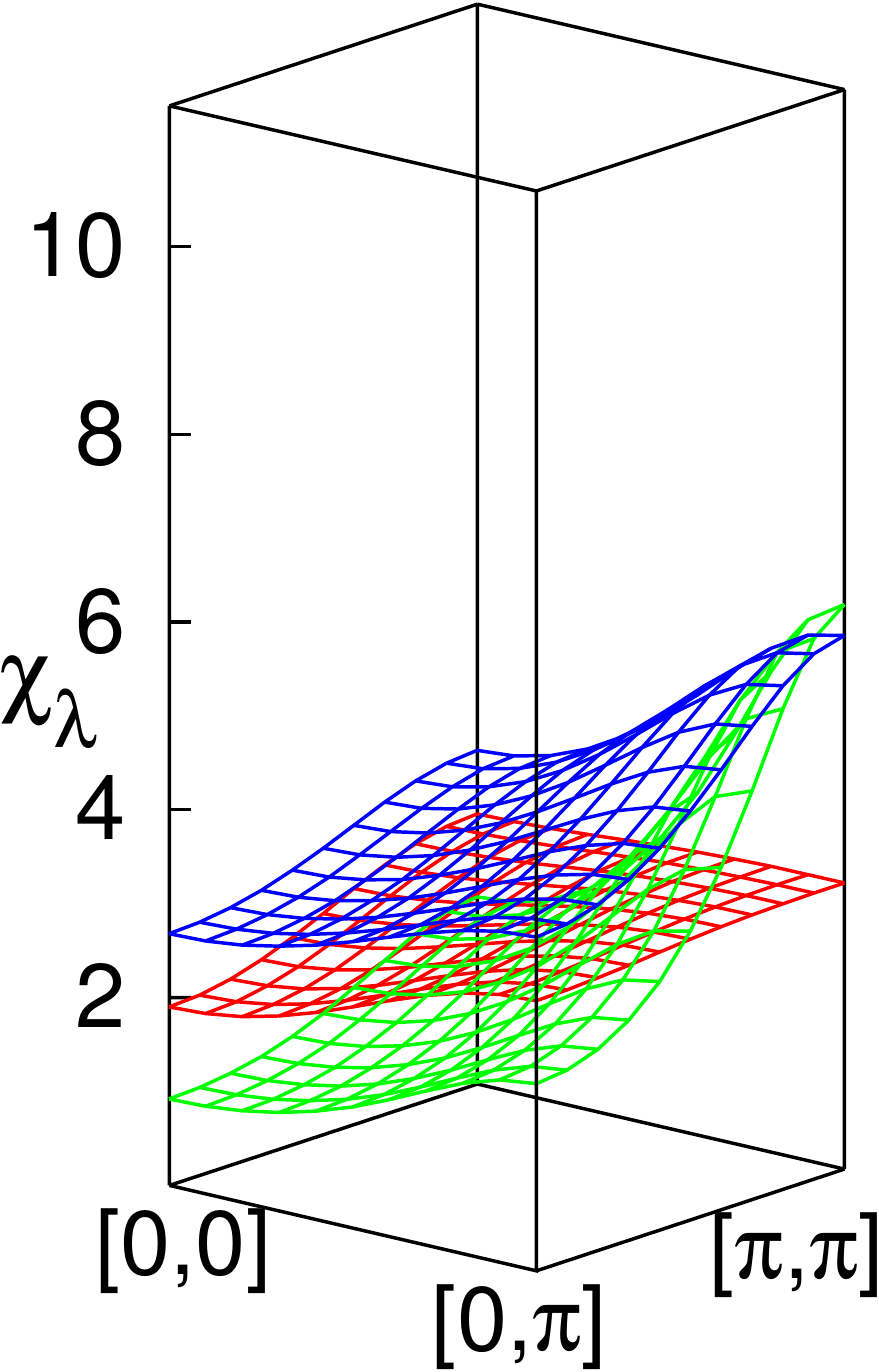}
      \includegraphics[width=0.15\columnwidth,clip]{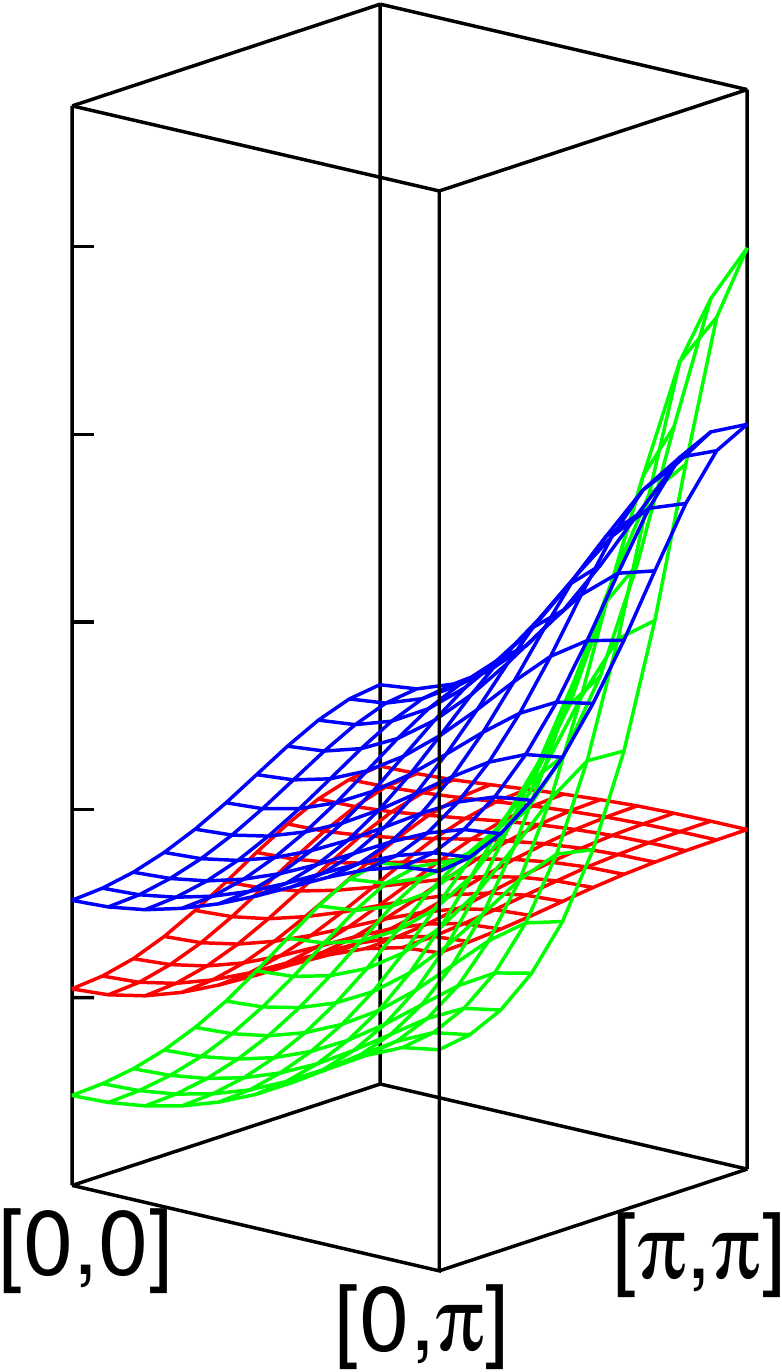}
      \includegraphics[width=0.15\columnwidth,clip]{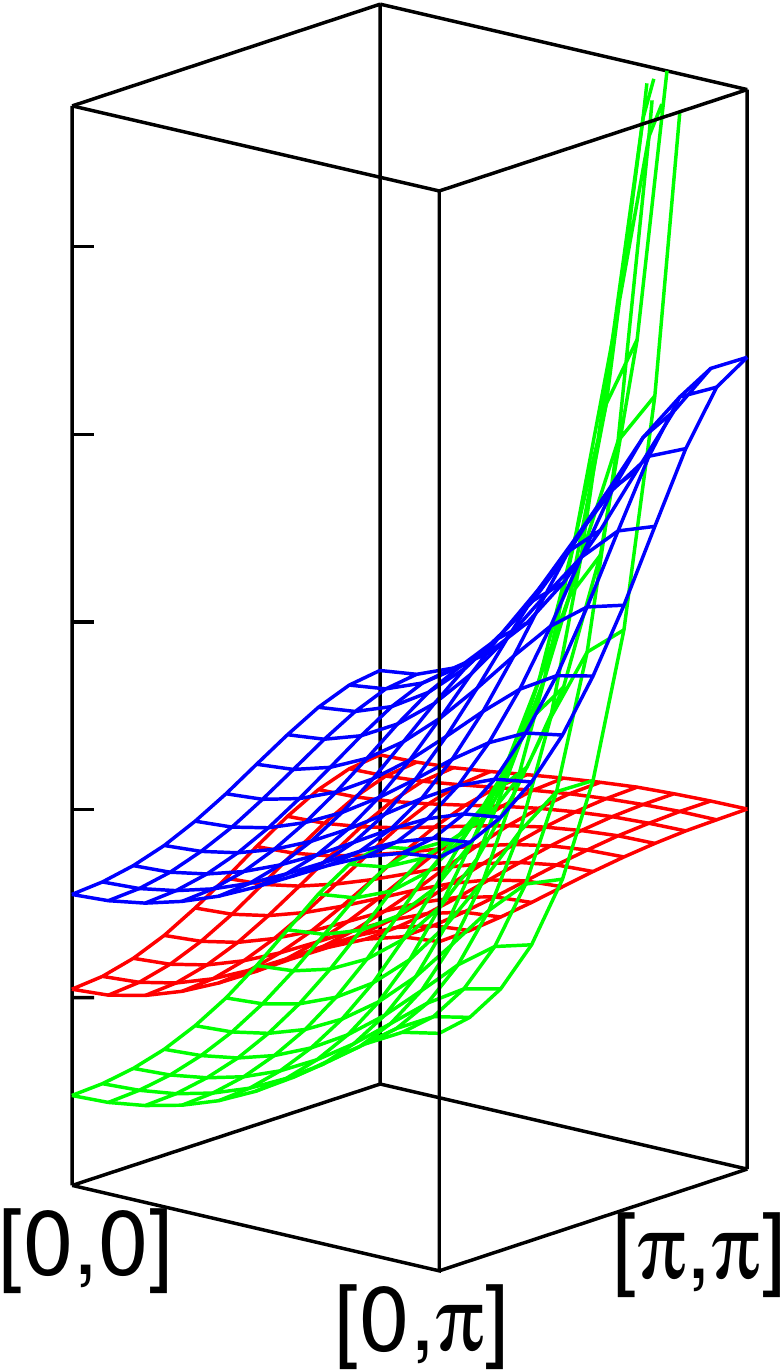}
      \includegraphics[width=0.4\columnwidth,clip]{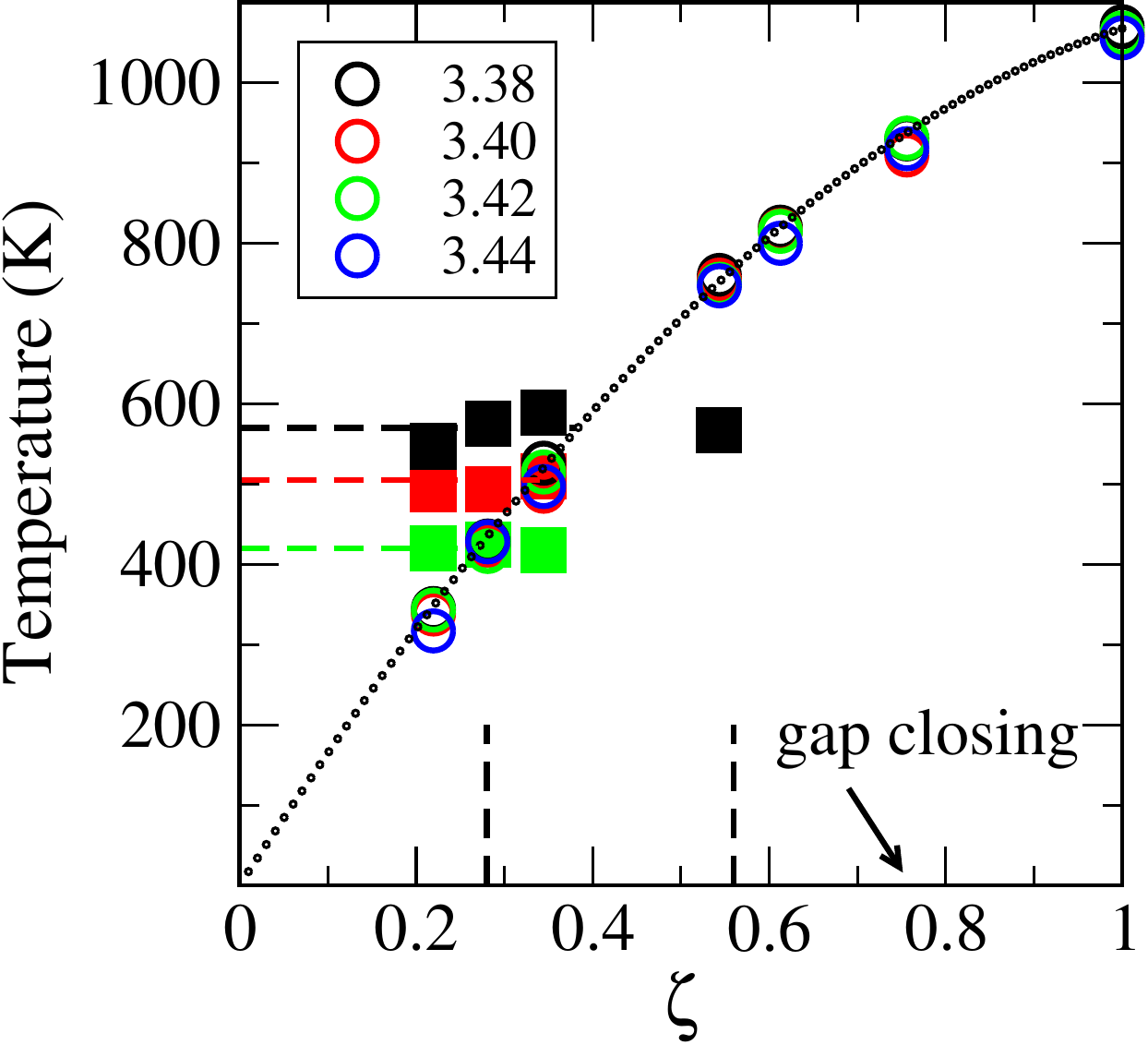}
    \end{center}
    \caption{ \label{fig:susc1} Left: The $\bq$-dependence of the leading eigenvalues of the particle-hole susceptibility matrix
      in a model with large band asymmetry $\zeta=\tfrac{2t_at_b}{t_a^2+t_b^2}=0.22$.
      The temperature decreases from left to right (773~K, 644~K, 580~K). The colors mark modes of the
      $s^z$ (red), $o$ (green) and degenerate $c^x$, $c^y$, $d^x$, $d^y$ (blue) character.
      Right: The instabilities towards exciton condensation ($c/d$ mode, open symbols) and spin-state ordering ($o$ mode, filled symbols) in the $\zeta-T$ plane for various values of the CF splitting $\Delta$. After Ref.~\cite{kunes14a}.}
  \end{figure}

  The linear-response formalism described in
  Sec.~\ref{sec:dmft-order} (Method 2) allows for an unbiased search for
  instabilities. Recently, several authors~\cite{kunes14a,boehnke15,hoshino16} applied
  this approach to the 2BHM in the parameter range close to the triple
  point between the metal and the HS and LS insulators. In Fig.~\ref{fig:susc1}
  we show a typical result of such a calculation. The generalized
  particle-hole susceptibility exhibits three distinct modes with
  sizeable amplitudes, which correspond to the spin susceptibility,
  orbital susceptibility, and spin-triplet excitonic
  susceptibilities, respectively~\cite{correspondence}. These
  describe the response of the system to the fields
  \begin{align}
    \label{eq:s}
    s_i^{\mu}&=\sum_{\alpha\beta}\left(a_{i\alpha}^{\dagger}\sigma^{\mu}_{\alpha\beta}a_{i\beta}^{\phantom\dagger}
      +b_{i\alpha}^{\dagger}\sigma^{\mu}_{\alpha\beta}b_{i\beta}^{\phantom\dagger}\right),\\
    \label{eq:o}
    o_i&=\sum_{\alpha}\left(a_{i\alpha}^{\dagger}a_{i\alpha}^{\phantom\dagger}
      -b_{i\alpha}^{\dagger}b_{i\alpha}^{\phantom\dagger}\right),\\
    \label{eq:d}
    d_i^{\mu}&=\sum_{\alpha\beta}\left(a_{i\alpha}^{\dagger}\sigma^{\mu}_{\alpha\beta}b_{i\beta}^{\phantom\dagger}
      +b_{i\alpha}^{\dagger}\sigma^{\mu}_{\alpha\beta}a_{i\beta}^{\phantom\dagger}\right),\\
    \label{eq:c}
    c_i^{\mu}&=-i\sum_{\alpha\beta}\left(a_{i\alpha}^{\dagger}\sigma^{\mu}_{\alpha\beta}b_{i\beta}^{\phantom\dagger}
      -b_{i\alpha}^{\dagger}\sigma^{\mu}_{\alpha\beta}a_{i\beta}^{\phantom\dagger}\right),
  \end{align}
  with site indices $i$, spin indices ${\alpha,\beta=\uparrow,\downarrow}$ and Cartesian indices $\mu=x,y,z$
  of the Pauli matrices $\sigma^{\mu}$.

  Note that in the actual calculations with the density-density
  interaction, it is only the spin $s_i^z$ and the excitonic
  $c_i^x,c_i^y,d_i^x,d_i^y$ components that lead to a finite
  response. The $U(1)$ spin rotation symmetry implies a degeneracy of
  the $x$ and $y$ excitonic modes. In the absence of cross-hopping
  (and pair-hopping) the spin-density $d_i^{\mu}$ and
  spin-current-density $c_i^{\mu}$ modes are degenerate. In
  Fig.~\ref{fig:susc1} both the orbital and the excitonic
  susceptibility diverge at $\bq=(\pi,\pi)$ when the temperature is
  decreased, with the orbital instability leading. This divergence is a
  signature of the checker board instability discussed in the previous
  section. Varying the ratio of the bandwidths one can switch between
  the orbital and excitonic instability. The orbital susceptibility is
  favored by strongly asymmetric bands, while for a moderate asymmetry
  the exciton condensation is the leading instability.
  \label{ssec:ec}

\subsubsection{Excitonic condensate}
  In the remainder of this section we study the properties of the exciton
  condensate (Method 1). We choose the model parameters in the
  region dominated by the excitonic instability (arrow in Fig.~\ref{fig:susc1}),
  but unlike in Fig.~\ref{fig:susc1} we choose, for computational convenience, $t_at_b<0$ to get a uniform
  condensate ($\bq=0$)~\cite{kunes15}.
    The condensate is characterized by a finite expectation value
    $\bph_i=\sum_{\alpha\beta} \boldsymbol{\sigma}_{\alpha\beta}\langle
  a_{i\alpha}^{\dagger}b_{i\beta}^{\phantom\dagger}\rangle=\langle \bd_i+i\mymathbf{c}_i\rangle$.
   It is instructive to consider the condensate wave function in
  the strong-coupling limit, which can be approximated as a product
  $\prod_i \left( s|\text{LS}\rangle_i+\xi|\text{HS}\rangle_i\right)$ of
   coherent superpositions of LS and HS states.
  Depending on the HS component we can obtain condensates with distinct symmetries and
  physical properties.
  In particular, HS states with  non-zero expectation value of the spin operator $\langle \bS_i \rangle\neq0$, e.g., $|S_z=1\rangle$
  give rise to a ferromagnetic condensate (FMEC) with finite, ferromagnetically ordered spin moments, while HS states with
  $\langle \bS_i \rangle=0$, e.g., $|S_z=0\rangle$ or $\tfrac{1}{\sqrt{2}}(|S_z=1\rangle+|S_z=-1\rangle)$,
  lead to a condensate with vanishing ordered moments which we call polar excitonic condensate (PEC).
  Which of these states is realized depends on the details of the model in Eq.~(\ref{eq:2bhm}).
  While FMEC and PEC states have many different properties they can be
  distinguished solely by the structure of the corresponding order parameter $\bph_i$~\cite{kunes14c}.

  \begin{figure}
    \includegraphics[width=0.9\columnwidth]{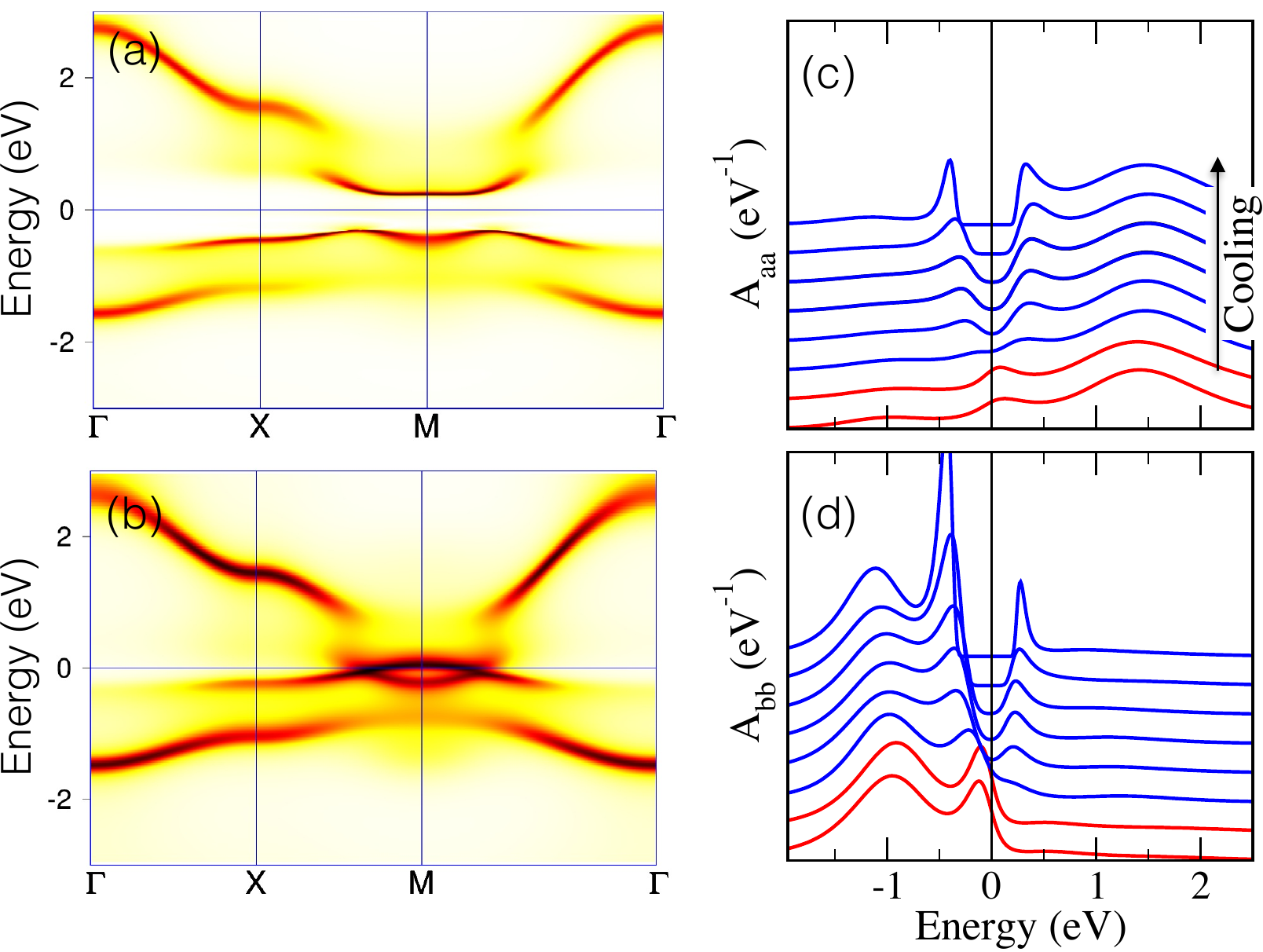}
    \caption{\label{fig:spec1} The $\bk$-resolved spectral function of the
      2BHM of Ref.~\cite{kunes14b} in (a) the exciton condensate (EC), and (b) the normal phase. (c),(d): Evolution of the local spectral for the same
      model between 1160~K and 290~K. The red and blue lines belong to
      the normal and EC phases, respectively. After
      Ref.~\cite{kunes14b}.}
  \end{figure}

  In addition to the bosonic physics of the condensate, the 2BHM describes
  also fermionic excitations and their interaction with the
  condensate.  In Fig.~\ref{fig:spec1} we show the evolution of the
  one-particle spectral function for the half-filled 2BHM when, starting from a
  semi-metallic normal state, the system is cooled through the EC
  instability. For the stoichiometric filling the system adopts the PEC
  phase, a preference that originates in the anti-ferromagnetic
  super-exchange between the HS states~\cite{kunes15}.
 Below the transition temperature a sizeable gap ($\gg
  k_BT$) opens. The similarity to an $s$-wave superconductor is not
  accidental. There is a mapping between the excitonic insulator and a
  superconductor, which dates back to the
  1960's~\cite{halperin68b,hoshino16} and involves a similar algebra for
  both phenomena. Another notable property of the PEC phase is a
  temperature independent spin susceptibility~\cite{kunes14b}.

\subsubsection{Doping and cross-hopping}
\begin{figure}
    \includegraphics[width=0.45\columnwidth,clip]{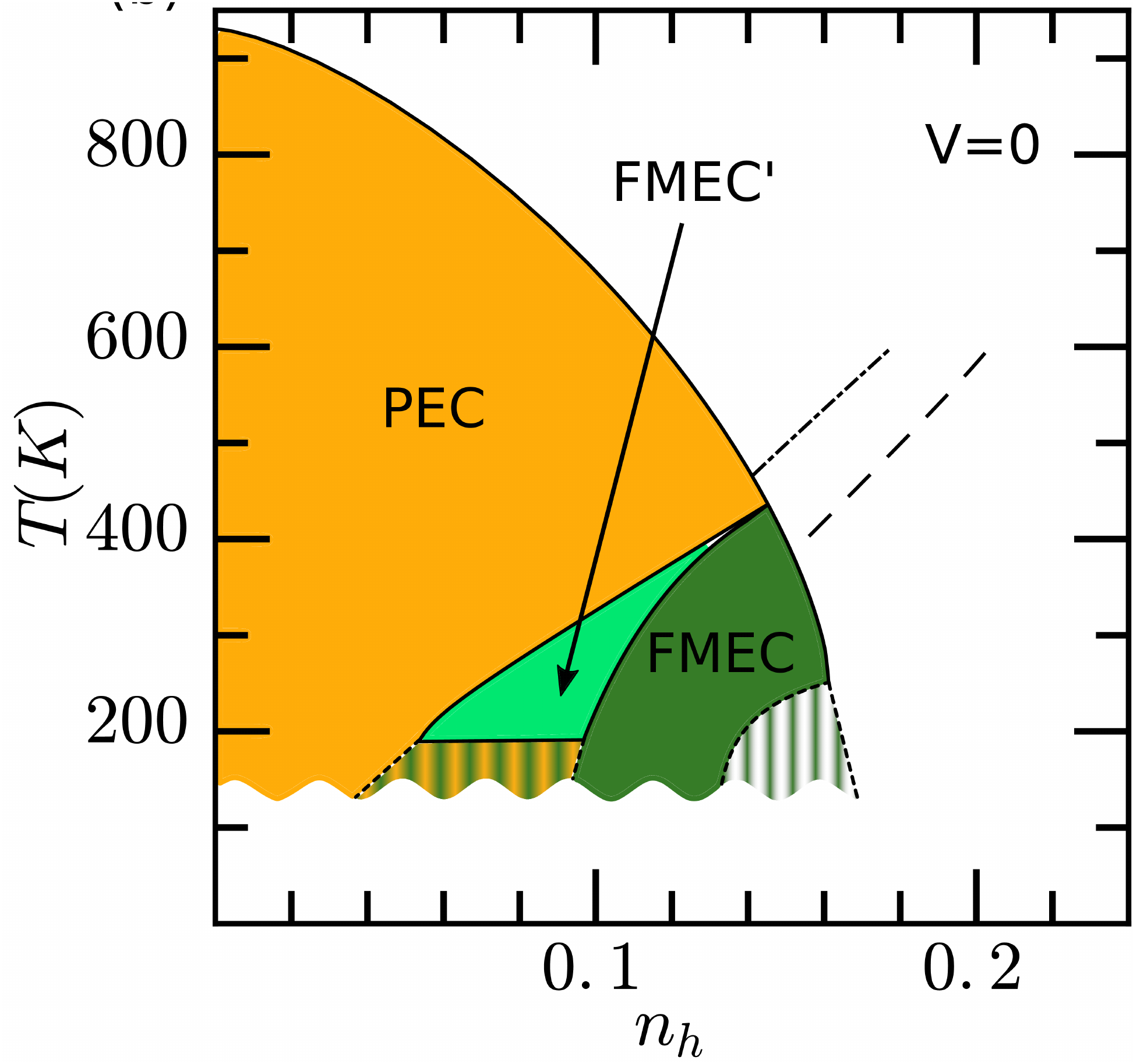}
    \includegraphics[width=0.38\columnwidth,clip]{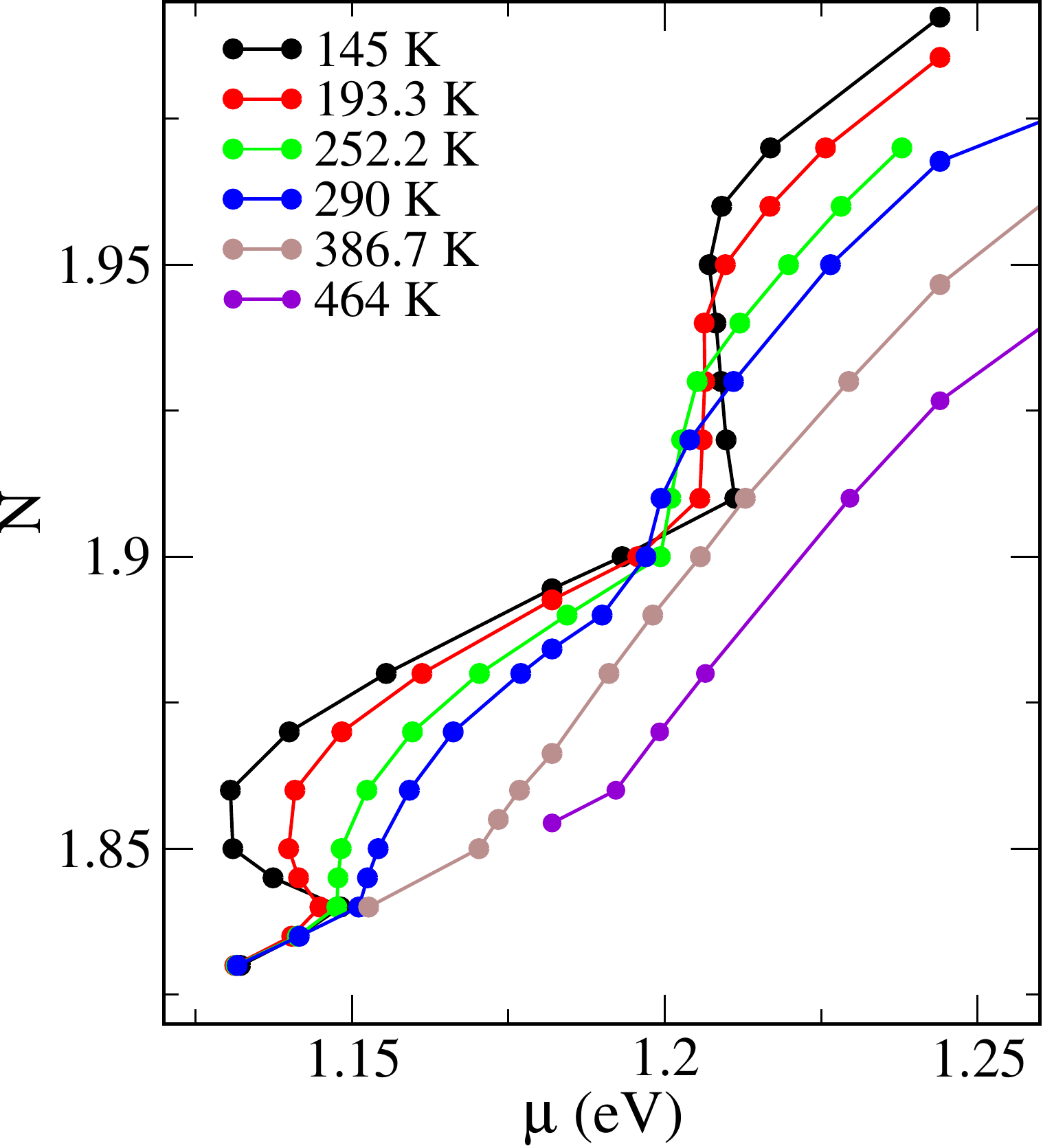}
    \caption{\label{fig:phs} Left: The phase diagram of the EC phase
      for the 2BHM without cross-hopping.  A description of the PEC, FMEC
      and FMEC$^\prime$ phases is given in the text.  Full lines denote
      continuous transitions, while dotted lines denote the boundaries of regions of
      phase coexistence.
      Right: Electron
      density vs. chemical potential along curves of constant temperature. The non-uniqueness of $N(\mu)$ indicates
      charge separation which can be quantified by a Maxwell
      construction.}
  \end{figure}

  Next, we consider the effect of doping on the exciton
  condensate. The DMFT phase diagram of a doped excitonic insulator is
  shown in Fig.~\ref{fig:phs}. A destructive effect of doping on the
  exciton condensation was observed experimentally~\cite{eisenstein04}
  and can be traced back to the competition between the condensation and the
  kinetic energy. In the present model the excitonic instability
  vanishes for more that 0.17 holes per atom. However, before that happens a
  transition to the FMEC phase takes place.  At low temperatures the
  transition is of first order and is accompanied by charge separation,
  while at higher temperatures it proceeds via two continuous phase
  transitions.  The microscopic origin of the FMEC phase has been explained in terms of
  a generalized double-exchange mechanism~\cite{kunes16}. The spin of the doped carrier couples to the net
  spin polarization of the condensate.  At low doping, the
  antiferromagnetic super-exchange between HS states prevails and the
  system settles in the PEC state. At higher doping, the spin polarized condensate
  is more transparent for the doped carriers and the gain in the kinetic
  energy stabilizes the FMEC state.

\begin{figure}
    \includegraphics[width=0.85\columnwidth,clip]{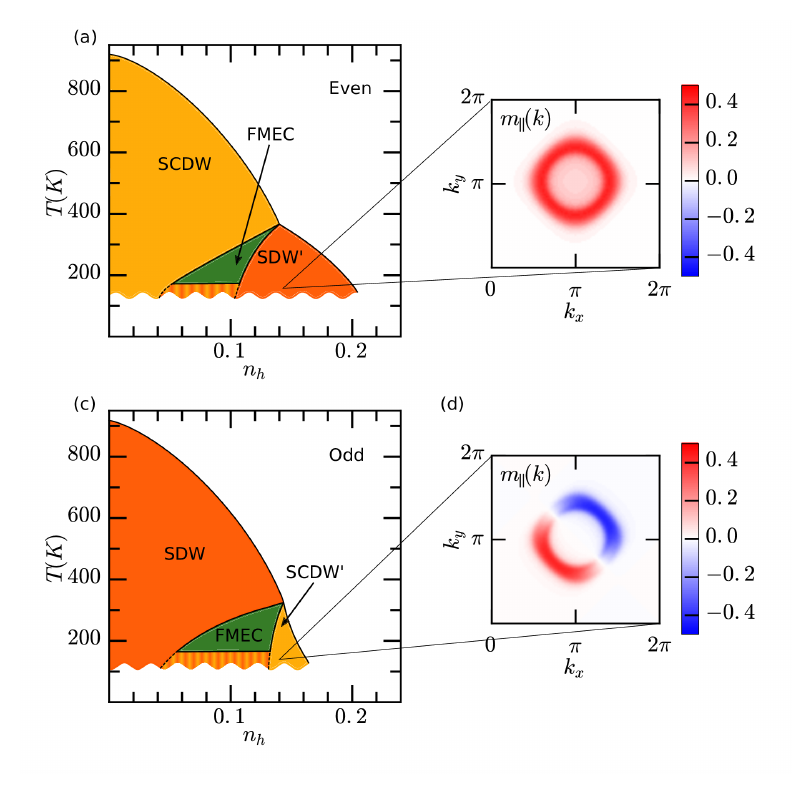}
    \caption{\label{fig:phase} Phase diagram calculated for the 2BHM with finite cross-hopping.  (a)
      and (c): Phase diagrams
      for even and odd cross-hopping, respectively.  Full lines denote
      continuous transitions, dotted lines denote the boundaries of
      phase coexistence regions.  (b) and (d): Spin textures in
      units of $\mu_B(a_0/2\pi)^2$ at the indicated points of the phase
      diagrams.  After Ref.~\cite{kunes16}.}
  \end{figure}

  Finally, we discuss the effect of moderate cross-hopping
  $|V_{ab}V_{ba}|\ll |t_at_b|$. We consider two
  possibilities for the cross-hopping: $V_{ab}=V_{ba}$ (even) and
  $V_{ab}=-V_{ba}$ (odd)~\cite{kunes16}. The corresponding phase
  diagrams, shown in Fig.~\ref{fig:phase}, exhibit the same basic
  features as their ``parent'' system in
  Fig.~\ref{fig:phs} with several notable differences.
  First, the PEC order parameter, which has an arbitrary
  complex phase (at least on the mean-field level) for $V_{ab}=0$, becomes purely real ($V_{ab}V_{ba}<0$) or imaginary ($V_{ab}V_{ba}>0$).
  The corresponding states are characterized by
  a local spin polarization with vanishing dipole but finite multipole
  moment (referred to as spin-density-wave (SDW) phase in the terminology of~\cite{halperin68b}), or by a local pattern of
  spin currents (referred to as spin-current-density-wave (SCDW) phase). Second, a continuous phase
  transition from the normal phase directly to the FMEC phase is possible only via an intermediate PEC phase
  (primed phases in Fig.~\ref{fig:phase}).

  All the SDW, SCDW, SDW$^\prime$, and SCDW$^\prime$ phases of Fig.~\ref{fig:phase}
  are instances of the PEC state. Nevertheless, the primed and
  unprimed phases have distinct properties.
  The primed phases were shown to host spin textures in the reciprocal
  space~\cite{kunes16}. In particular, textures with odd $\bk$-parity are interesting
  since they do not break time reversal symmetry and give rise to
  an effective Hamiltonian similar to Rashba-Dresselhaus spin-orbit
  coupling. Microscopically the spin textures can be traced to an
   effective spin dependent hopping of the doped carriers on the condensate background~\cite{kunes16}.

\section{Structural stability of correlated materials}
\label{sec:lattice}

  In this section we present a detailed formulation of a variant of the DFT+DMFT
  computational scheme~\cite{metzner89,georges96,Kotliar04,Kotliar06,Anisimov97,Lichtenstein98,Held06},
  which allows one to compute atomic displacements and thereby detect
  correlation-induced structural transformations, determine the phase stability, calculate the
  lattice dynamics, and perform a structural optimization of
  correlated electron materials~\cite{Leonov08,Leonov08_,Leonov08__,Leonov11,Leonov12,Leonov15a}.

  We also present a detailed formulation of a fully charge self-consistent DFT+DMFT scheme,
  which includes the effect of electron correlations on the charge
  density~\cite{Pourovskii07,Haule07,Amadon08,Aichhorn09,Park14}.
  The correlations
  lead to a charge redistribution between the correlated $3d$ or $4f$
  orbitals and non-correlated $s,p$ states, which has been shown to
  play an important role, e.g., near the Mott metal-insulator
  transition~\cite{Leonov15a}.

The DFT+DMFT approach
overcomes the limitations of standard band-structure
  techniques and allows one to perform microscopic investigations of
  the electronic structure and lattice properties of correlated
  electron materials.
  It has been widely employed in recent
  investigations of the electronic and structural properties of
  strongly correlated electron
  materials~\cite{Savrasov01,Savrasov01_,McMahan03,Amadon06,Bieder14,lichtenstein01,Katanin10,Leonov11,Leonov12,Belozerov13,Leonov14a,Glazyrin13,Glazyrin13_,Held01,Held01_,Held01__,Held01___,Poteryaev07,Grieger12,Grieger12_,leonov15,Leonov15a,Leonov15c,Pavarini04,Biermann05,kunes09,Poteryaev04,Leonov08,Leonov08_,Leonov08__,Leonov14b,Pavarini08,Pavarini10,kunes08,Kunes08_,shorikov10,ohta12,Huang12,Huang12_,Ren06,Ren06_,Ren06__,Ren06___,Ren06____,Byczuk12,Byczuk12_}. %

  In particular, this method was used to study
  the electronic
  and structural properties of elemental
  Fe~\cite{Leonov11,Leonov12,Leonov14a} and the iron chalcogenide FeSe \cite{Leonov15c}, and to explain the cooperative Jahn-Teller effect in
  paramagnetic KCuF$_3$ and LaMnO$_3$~\cite{Leonov08,Leonov08_,Leonov08__}. Recently it was employed to explore the electronic properties and the
  phase stability of paramagnetic V$_2$O$_3$~\cite{Leonov15a} and
  the transition metal monoxides MnO, FeO, CoO, and NiO near the
  pressure-induced Mott-Hubbard metal-insulator
  transition~\cite{leonov16}.

  We will demonstrate that the DFT+DMFT approach provides a qualitative and
  quantitative description of the electronic properties and phase
  stability of all these materials, in spite of their chemical,
  structural, and electronic differences. The scheme is robust and
  makes it possible to address, on the same footing, electronic,
  magnetic, and structural properties of strongly correlated
  materials.

\subsection{Methodological developments: Total-energy functional and a fully charge self-consistent DFT+DMFT scheme}
\label{sec:method}

  In order to compute the electronic properties and determine the phase stability of
  correlated electron materials, we implemented~\cite{Leonov15a}
  a fully charge self-consistent DFT+DMFT
  approach~\cite{Pourovskii07,Haule07,Amadon08,Aichhorn09,Park14},
  which is formulated in terms of plane-wave
  pseudopotentials~\cite{Baroni01,Giannozzi09}. Following Anisimov \emph{et al.}~\cite{Anisimov05} and Trimarchi \emph{et
    al.}~\cite{Trimarchi08} we construct a basis set of
  atomic-centered symmetry-constrained Wannier
  functions~\cite{Marzari97,Anisimov05,Trimarchi08,Korotin08}, which
  allows us to build up an effective low-energy Hamiltonian
  $\hat{H}_{\text{DFT}}$ for the partially occupied correlated orbitals,
  e.g., $d$ of $f$ orbitals of transition-metal ions. The Hamiltonian
  $\hat{H}_{\text{DFT}}$, which provides a realistic description of the
  single-electron band structure of a material, is further
  supplemented by on-site Coulomb interactions for the correlated
  orbitals. In the
  density-density approximation for the electronic interaction (see Sec.~\ref{sec:2bhm})
  the  many-body Hubbard Hamiltonian  takes the form
  \begin{eqnarray}
    {\hat H} & = & {\hat H_{\text{DFT}}} + \frac{1}{2} \sum_{m,\sigma}
    U^{\sigma\sigma'}_{mm'} \hat n_{m \sigma} \hat n_{m'\sigma'} - {\hat H_{\text{DC}}},
    \label{eq:hamiltonian}
  \end{eqnarray}
  where $\hat n_{m \sigma}= \hat c^\dagger_{m \sigma} \hat c_{m \sigma}$
  is the local density operator for the orbital $m$ and spin $\sigma$.
  Here $U^{\sigma,\sigma}_{m,m'}$ and
  $U^{\sigma,\bar{\sigma}}_{m,m'}$ are the reduced interaction
  matrices for equal and opposite spins, respectively. The interaction
  matrices are expressed in terms of the Slater integrals $F^0$,
  $F^2$, and $F^4$, which for the case of $3d$ electrons are related
  to the local Coulomb and Hund's rule coupling as $U=F^0$,
  $J=(F^2+F^4)/14$, and $F^2/F^4 = 0.625$. The term $\hat H_{\text{DC}}$ is a
  double-counting correction which accounts for the electronic
  interactions already described by DFT. The Coulomb repulsion $U$ and
  Hund's coupling $J$ can be evaluated using constrained DFT and/or
  constrained random phase approximation (RPA) methods within a
  Wannier-functions formalism~\cite{Korotin08}. The $U$ and $J$ values
  are related to the matrix notation as $U = \sum_{\{m\}}
  U^{\sigma\bar{\sigma}}_{mm'} / N^2$ and $J=U- \sum_{\{m \neq m'\}}
  U^{\sigma\sigma}_{mm'}/ N(N-1)$, where $N$ is the number of correlated
  orbitals.

  We implement the fully charge self-consistent DFT+DMFT approach in
  terms of plane-wave ultrasoft
  pseudopotentials~\cite{Baroni01,Giannozzi09}. In this scheme, the
  matrix elements of $\hat H_{\text{DFT}}$ are determined as
  \begin{eqnarray}
    [H_{\text{DFT}}^\sigma(\mymathbf{k})]_{\mu \nu} = \sum_{i} P^{\sigma*}_{i\mu}(\mymathbf{k}) \epsilon^{\sigma}_{\mymathbf{k}i} P^{\sigma}_{i\nu}(\mymathbf{k}),
    \label{eq:hdft}
  \end{eqnarray}
  where $P^{\sigma}_{i\nu}(\mymathbf{k}) = \langle
  \psi^{\sigma}_{\mymathbf{k}i}|\hat{S}| \phi^{\sigma}_{\mymathbf{k}\nu}
  \rangle$ are the matrix elements of the orthonormal projection
  operators expressed in the basis of local orbitals
  $\phi^{\sigma}_{\mymathbf{k}\nu}$.  The charge density
  $\rho(\mymathbf{r})$ is calculated as
  \begin{eqnarray}
    \rho( \mymathbf{r} )= T \sum_{\mymathbf{k},i\omega_n;ij} \rho_{ \mymathbf{k};ij } G_{\mymathbf{k};ji}( i\omega_n ) e^{i\omega_n0+},
    \label{eq:rho}
  \end{eqnarray}
  where $T$ is the temperature. The charge density matrix elements
  $\rho_{ \mymathbf{k};ij }$ in the basis of the Kohn-Sham wave
  functions $\psi_{\mymathbf{k}i}$ are computed as
  \begin{eqnarray}
    \rho_{\mymathbf{k};ij}( \mymathbf{r} )= \langle \psi_{\mymathbf{k}i}| \mymathbf{r} \rangle \langle \mymathbf{r}|\psi_{\mymathbf{k}j} \rangle + \sum_{I,lm} Q_{lm}({\bm{r} - \bm{R}_I}) \langle \psi_{\mymathbf{k}i}| {\beta^I_{l}} \rangle \langle {\beta^I_{m}} | \psi_{\mymathbf{k}j} \rangle.
    \label{eq:chrage_density}
  \end{eqnarray}
  Here, the index $I$ refers to an atomic site, the functions
  $\beta^I_{m}( \mymathbf{r} )$ are computed in an atomic calculation
  and differ for different atomic species. Furthermore, $Q^I_{lm}(\mymathbf{r})$ is
  the augmentation function that is strictly localized in the core
  region, and $G_{\mymathbf{k};ij} ( i\omega_n )$ is the lattice Green's
  function in the KS wave functions basis at a given $\bm{k}$-point. The (inverse) lattice Green's function is evaluated as
  \begin{eqnarray}
    G^{\sigma}(\mymathbf{k},i\omega_n)^{-1}_{ij} = (i\omega_n + \mu - \epsilon^{\sigma}_{\mymathbf{k}i})\delta_{ij}-\Sigma^{\sigma}_{ij}(\mymathbf{k},i\omega_n),
    \label{eq:lattice_gf}
  \end{eqnarray}
  where $\epsilon^{\sigma}_{\mymathbf{k}i}$ are the Kohn-Sham
  eigenvalues. $\Sigma^{\sigma}_{ij}(\mymathbf{k},i\omega_n)$ is the
  self-energy obtained from the solution of the DMFT impurity problem
  by ``upfolding" of the impurity self-energy from the localized
  Wannier to the Kohn-Sham basis. We note that for the non-correlated states
  which do not hybridize with the correlated orbitals the density
  matrix collapses to the Fermi function for the state $\mymathbf{k}$ and
  band $i$, i.e., $T\sum_{i\omega_n}G_{\mymathbf{k};ij} ( i\omega_n
  )e^{i\omega_n0+} = f_{\mymathbf{k}i}\delta_{ij}$. In practice, it is
  convenient to compute the charge density as $\rho( \bm{r} ) =
  \rho_{\text{DFT}}( \mymathbf{r} ) + \Delta \rho( \mymathbf{r} )$, i.e., to
  split the DFT contribution and the correlation-induced difference
  $\Delta \rho( \mymathbf{r} )$. Full charge self-consistency is
  achieved when both the charge density $\rho(\mymathbf{r})$ and the local
  Green's function are converged.

  Within the DFT+DMFT calculation  the total energy is then evaluated using the expression~\cite{Amadon06,Leonov08,Leonov08_,Leonov08__}
  \begin{equation}
    E = E_{\text{DFT}}[\rho( \mymathbf{r} )] + \langle \hat H_{\text{DFT}} \rangle - \sum_{m,\mymathbf{k}}
    \epsilon^{\text{DFT}}_{m\mymathbf{k}} + \langle \hat H_{U} \rangle - E_{\text{DC}},
    \label{eq:energy}
  \end{equation}
  where $E_{\text{DFT}}[\rho (\mymathbf{r}) ]$ is the DFT total energy obtained
  for the self-consistent charge density $\rho( \mymathbf{r} )$. The
  third term on the right-hand side of Eq.~\eqref{eq:energy} is the sum
  of the DFT valence-state eigenvalues which is evaluated as the
  thermal average of the DFT Hamiltonian with the non-interacting DFT
  Green's function $G^{\text{DFT}}_{\mymathbf{k}}(i\omega_n)$:
  \begin{equation}
    \sum_{m,\mymathbf{k}} \epsilon^{\text{DFT}}_{m,\mymathbf{k}} = T\sum_{i\omega_n,\mymathbf{k}}
    \Tr[H_{\text{DFT}}(\mymathbf{k}) G^{\text{DFT}}_{\mymathbf{k}}(i\omega_n)] e^{i\omega_n0^{+}}.
  \end{equation}
  $\langle \hat H_{\text{DFT}} \rangle$ is evaluated in a similar way but
  with the full Green's function including the self-energy. To
  calculate these two contributions, the summation is performed over
  the Matsubara frequencies $i\omega_n$, taking into account an
  analytically evaluated asymptotic correction. Thus, for $\langle
  \hat H_{\text{DFT}} \rangle$ one has
  \begin{eqnarray}
    \langle \hat H_{\text{DFT}} \rangle &=& T\sum_{i\omega_n,\mymathbf{k}} \Tr[H_{\text{DFT}}(
    \mymathbf{k}) G_{\mymathbf{k}}(i\omega_n)] e^{i\omega_n0^{+}}  \notag \\
    &=& T\sum_{i\omega_n,\mymathbf{k}} \Tr \{ H_{\text{DFT}}(
    \mymathbf{k}) [ G_{\mymathbf{k}}(i\omega_n)- \frac{m^{\mymathbf{k}}_{1}}{%
      (i\omega_n)^2} ] \}  \notag \\
    &&+\,\frac{1}{2} \sum_{\mymathbf{k}} \Tr[H_{\text{DFT}}(\mymathbf{k})]- \frac{1}{4T}
    \sum_{\mymathbf{k}} \Tr[H_{\text{DFT}}(\mymathbf{k}) m^{\mymathbf{k}}_{1}]
  \end{eqnarray}
  where the first moment $m^{\mymathbf{k}}_{1}$ is computed as
  $m^{\mymathbf{k}}_{1} = H_{\text{DFT}}(\mymathbf{k}) + \Sigma(i\infty)- \mu$.
The asymptotic part of the self-energy $\Sigma(i \infty)$ is
  calculated as the average of $\Sigma(i\omega_n)$ over the last
  several $i\omega_n$ points.  The interaction energy $\langle \hat
  H_{U} \rangle$ is computed from the double occupancy matrix. The
  double-counting correction $E_{\text{DC}}$ is evaluated as the average
  Coulomb repulsion between the $N_{d}$ correlated electrons in the
  Wannier orbitals.

  This DFT+DMFT approach allows us to determine correlation-induced
  structural transformations of correlated materials together with the corresponding change of
  the atomic coordinates and the unit cell shape. It can also be
  used to explain experimentally observed structural data
  and to predict structural properties of real correlated
  materials.

In particular, we performed fully charge self-consistent computations
  of the
  electronic structure and phase stability of
  paramagnetic V$_2$O$_3$ across the Mott-Hubbard metal-insulator
  transition (MIT)~\cite{Leonov15a}; see also Refs.~\cite{Grieger12,Grieger12_}.
  Namely, in charge \emph{non}-self-consistent calculations the $a_{1g}$--$e_g^{\pi}$ crystal-field splitting of paramagnetic V$_2$O$_3$
 is found to be strongly enhanced~\cite{Held01,Held01_,Held01__,Held01___,Poteryaev07}. This leads to a substantial redistribution of the charge density and
thereby influences the lattice structure due to electron-lattice coupling. For that reason
full charge self-consistency turns out to be crucial to obtain a more realistic description of the physical properties of
  V$_2$O$_3$~\cite{Leonov15a} near the
  Mott metal-insulator transition. For a detailed discussion we refer to Ref.~\cite{Leonov15a}.

  In the following we report our results obtained by the
DFT+DMFT scheme
  for the electronic and structural properties of elemental
  Fe~\cite{Leonov11,Leonov12,Leonov14a} and the iron chalcogenide FeSe \cite{Leonov15c}.

\subsection{Lattice dynamical properties of paramagnetic Fe}
\label{sec:iron}

  Elemental iron is an exceptionally important material even for
  present-day technology. Iron exhibits a rich phase
  diagram with at least four allotropic forms~\cite{Basinski55,Wyckoff63}. At
  ambient conditions it is ferromagnetic and has a bcc crystal
  structure ($\alpha$ iron). Upon heating above the Curie temperature $T_C\sim$ 1043 K,  $\alpha$
  iron becomes paramagnetic, but remains in its bcc
  crystal structure. Only upon further increase of the temperature above $T_{\text{struct}} \sim$
  1185 K $\alpha$
  iron exhibits a structural phase transition to a fcc structure
  ($\gamma$ phase). Under pressure $\alpha$ iron makes a transition to
  a paramagnetic hcp structure ($\epsilon$ phase) at $\sim$ 11 GPa.

\subsubsection{Lattice stability and phonon spectra near the $\alpha$-$\gamma$ structural transition}

  State-of-the-art band structure methods provide a qualitatively
  correct description of various electronic and structural properties
  of
  iron~\cite{Hasegawa83,Singh91,Singh91_,Singh91__,DalCorso00,Singh91____,Sha06,Kormann12,Ruban12}. For
  example, these methods provide a good quantitative understanding of
  the equilibrium crystal structure and the lattice dynamical
  properties of the ferromagnetic $\alpha$ phase. However,
  applications of these techniques to describe, e.g., the
  $\alpha$-$\gamma$ phase transition in iron, do not lead to
  satisfactory results. They predict a simultaneous transition of the
  structure and the magnetic state at the bcc-fcc phase transition
  while, in fact, the bcc-to-fcc phase transition occurs only about
  150 K above $T_C$. Moreover, the elastic and dynamical stability of
  the bcc phase is found to depend sensitively on the value of the
  magnetization. For example, in the absence of the magnetization,
  standard band-structure methods predict bcc iron to be
  unstable~\cite{Hsueh02}. We now understand that this is due to the
  presence of local moments above $T_C$ which cannot be treated realistically
  by conventional band structure techniques.

  This problem has been overcome by employing the DFT+DMFT approach
  which allows one to study correlated materials both in to the
  long-range ordered and paramagnetic
  state~\cite{lichtenstein01,Katanin10,Leonov11,Leonov12,Belozerov13,Leonov14a,Glazyrin13,Glazyrin13_}. The
  DFT+DMFT method naturally accounts for the existence of local moments
  above $T_C$ and has shown to provide a good quantitative
  description of the properties of $\alpha$ iron. Moreover,
  applications of DFT+DMFT to study the equilibrium crystal structure
  and phase stability of iron at the $\alpha$-$\gamma$ phase
  transition reveal that the bcc-to-fcc phase transition takes place at a temperature of about 1.3 $T_C$,
  i.e., well above the magnetic transition, in
  agreement with experiment~\cite{Leonov11}.

  We determine the structural phase stability and lattice dynamics of
  paramagnetic iron at finite temperatures by employing the DFT+DMFT
  approach implemented with the frozen-phonon
  method~\cite{Leonov12,Leonov14a}. The approach is implemented with
  plane-wave pseudopotentials~\cite{Baroni01,Giannozzi09} which allows
  us to compute lattice transformation effects caused by electronic
  correlations~\cite{Leonov08,Leonov08_,Leonov08__}. We employ this technique to study the
  temperature dependent phonon dispersion relations and phonon spectra
  of paramagnetic iron at the bcc-fcc phase transition. Our presentation follows the discussion of Ref.~\cite{Leonov12}.

  It was previously shown that band structure calculations within the nonmagnetic generalized gradient approximation
  (GGA) cannot explain the experimentally observed phase
  stability of paramagnetic iron at the bcc-fcc phase transition,
  since they do not describe electronic correlations adequately.
  We now include the effect of electronic correlations by constructing
  an effective low-energy Hamiltonian for the partially filled Fe $s$, $d$
  orbitals based on the results of the nonmagnetic GGA. We construct a
  basis of atomic-centered symmetry-constrained Wannier functions for
  the Fe $s$, $d$ orbitals~\cite{Anisimov05,Trimarchi08,Korotin08}, with
  $U = 1.8$ eV and $J = 0.9$ eV  as obtained by
  previous theoretical and experimental estimations. To solve the
  realistic many-body problem we employ the Hirsch-Fye algorithm without charge self-consistency.

The phase stability and lattice dynamical properties of
  iron near the bcc-fcc phase transition are computed within the DFT+DMFT
  approach implemented with the frozen-phonon
  method~\cite{Leonov12}. The phonon frequencies are calculated by
  introducing a small set of displacements in the corresponding
  supercells of the equilibrium lattice which results in an energy difference
  with respect to the undistorted structure. We
  first focus on the lattice dynamical properties of iron near the
  bcc-to-fcc phase transition and perform calculations at temperatures
  $T = 1.2~T_C$ and 1.4 $T_C$, which are below and above the
  temperature $T_{\text{struct}} \sim 1.3~T_C$ where the structural phase
  transition occurs.

  \begin{figure}[tbp!]
    \centerline{
      \includegraphics[width=0.5\textwidth,clip=true]{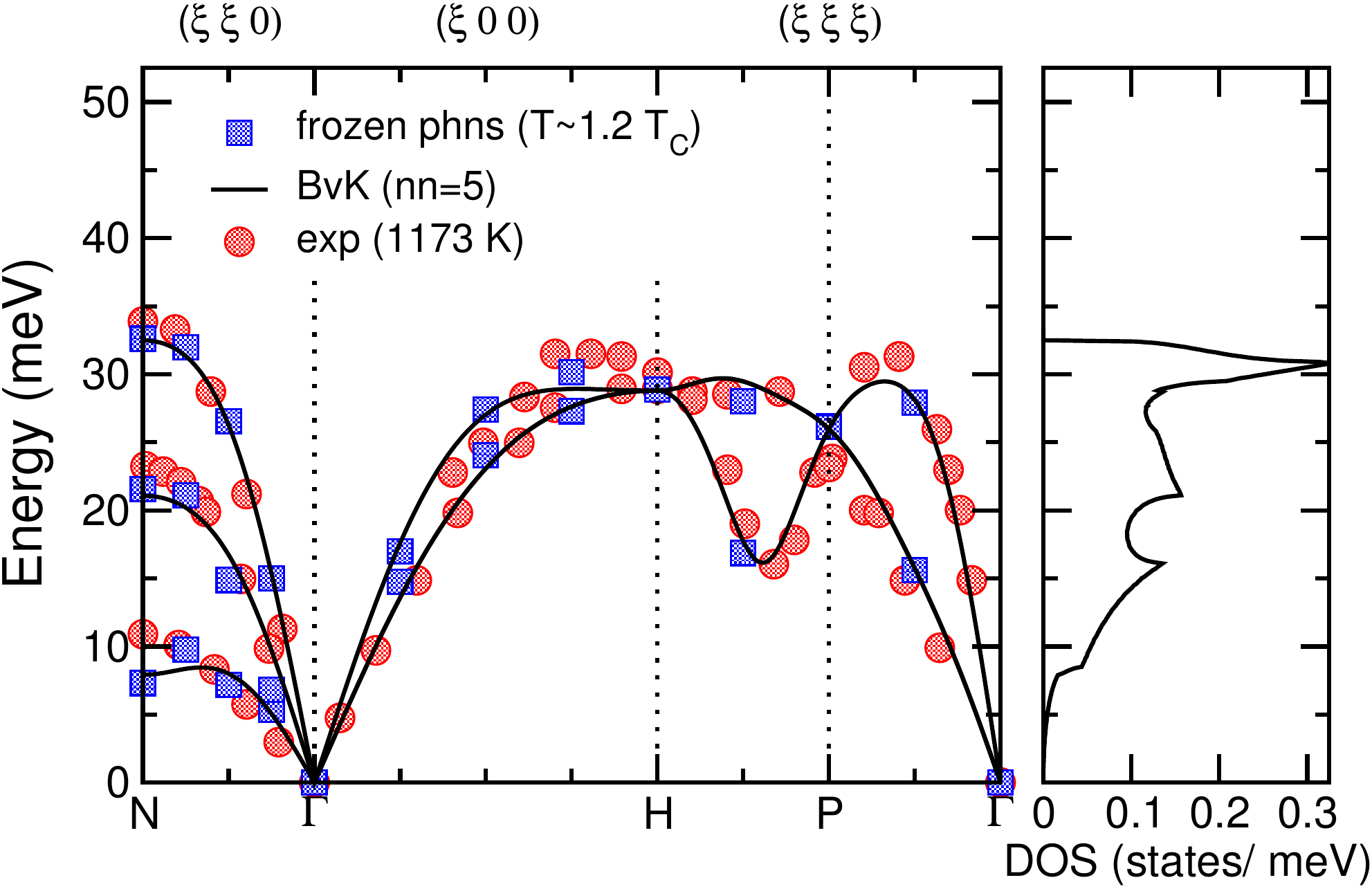}
      \includegraphics[width=0.5\textwidth,clip=true]{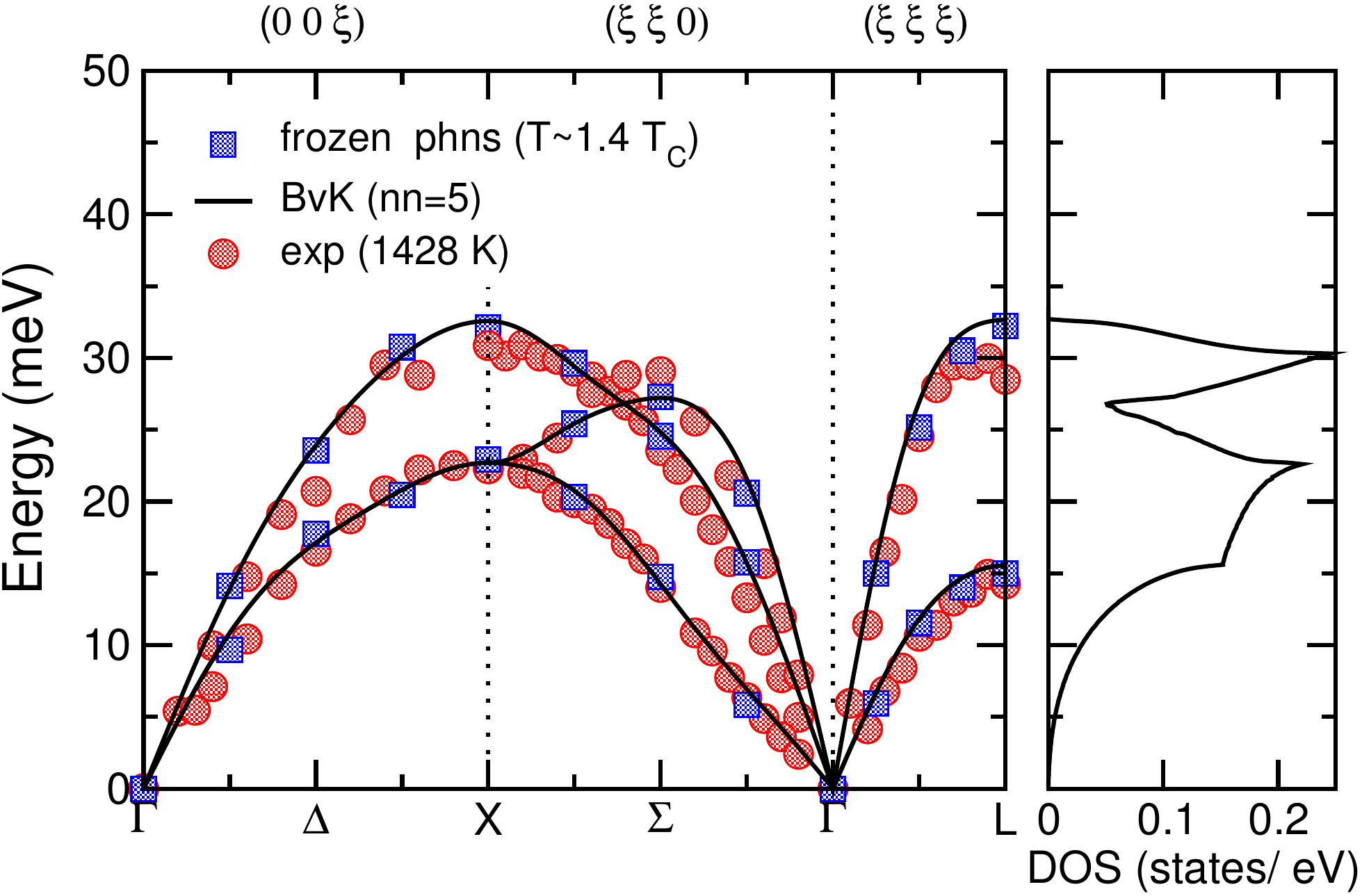}}
    \caption{Phonon dispersion relations and corresponding phonon
      density of states (DOS) of paramagnetic bcc (left) and fcc Fe (right)
      as calculated within DFT+DMFT \cite{Leonov12}. The DFT+DMFT result is further
      interpolated using a Born-von K\' arm\' an model with
      interactions expanded up to the 5-th nearest-neighbor shell. The
      results are compared with neutron inelastic scattering
      measurements at 1173 K \cite{Neuhaus97}  and 1428 K \cite{Zarestky87} for the bcc and the fcc Fe
      phase, respectively. From Ref.~\cite{Leonov12} with permission of the authors.}
    \label{Fe:Fig_1}
  \end{figure}

  We present our results for the phonon dispersion relations and
  phonon spectra in Fig.~\ref{Fe:Fig_1}. The computations are
  performed for the   equilibrium volume calculated at this particular temperature
 (lattice constant $a$ = 2.883 \AA{},  which almost coincides with experiment). To evaluate the phonon
  frequencies for arbitrary wave vectors in the Brillouin zone, we
  performed lattice dynamical calculations on the basis of a Born-von
  K\' arm\' an model with interactions expanded up to the 5-th
  nearest-neighbor shell. The calculated phonon dispersions of the bcc
  phase of iron show the typical behavior of a bcc metal with an
  effective Debye temperature $\sim$ 458 K. The phonon frequencies are
  overall positive, implying mechanical stability of the bcc lattice
  structure at $\sim$ 1.2 $T_C$, i.e., well above the Curie
  temperature, in agreement with experiment. This corrects the results obtained
  with the non-magnetic GGA which finds the bcc lattice to be
  dynamically unstable even for the equilibrium lattice constant $a$ $=$
  2.883 \AA{}. Most
  importantly, our calculations clearly demonstrate the crucial
  importance of electronic correlations to explain both the
  thermodynamic and the lattice dynamical stability of the
  paramagnetic bcc phase of iron. For details we refer to Ref.~\cite{Leonov12}. Overall, the structural phase
  stability, equilibrium lattice constant, and phonon frequencies of
  bcc iron obtained by DFT+DMFT are in remarkably good agreement with
  the experimental data which were taken at nearly the same reduced
  temperature $T/T_C$~\cite{Neuhaus97}.

  We also study the lattice dynamical properties of the
  paramagnetic fcc phase which is found to become energetically
  favorable above $\sim 1.3~T_C$. To prove the mechanical stability of
  the fcc phase we evaluate the lattice dynamics of the fcc phase at
  $T \sim 1.4~T_C$. Our results for the phonon dispersion
  relations and phonon spectra, which were calculated for the
  equilibrium lattice constant $a = 3.605$ \AA{} are also shown in
  Fig.~\ref{Fe:Fig_1}. The effective Debye temperature is obtained as $\sim$ 349 K. The phonon frequencies are overall positive,
  implying mechanical stability of the fcc lattice structure at $T
  \sim 1.4~T_C$. Although
  nonmagnetic GGA calculations also find the fcc lattice structure to be mechanically stable (for the GGA equilibrium volume),
  the GGA energy for fcc iron is \emph{higher} than that for the
  close-packed hcp structure. By contrast, DFT+DMFT calculation
  find the simultaneous thermodynamic and lattice dynamical stability
  of the paramagnetic fcc phase of iron at $\sim 1.4~T_C$, in
  agreement with experiment. Altogether our results for the structural phase
  stability, equilibrium lattice constant, and phonon frequencies
  agree remarkably well with the available experimental data taken at
  nearly the same reduced temperature $T \sim
  1.4~T_C$~\cite{Zarestky87}. Again it should be noted that the
  application of the nonmagnetic GGA to fcc iron finds phonon
  frequencies which differ considerably from experiment. These
  findings clearly demonstrate the importance of electronic
  correlations for the lattice dynamical properties of fcc iron.

\subsubsection{Origin of the phase stability of $\delta$ iron}

\begin{figure}[tbp!]
    \centerline{
      \includegraphics[width=0.5\textwidth,clip=true]{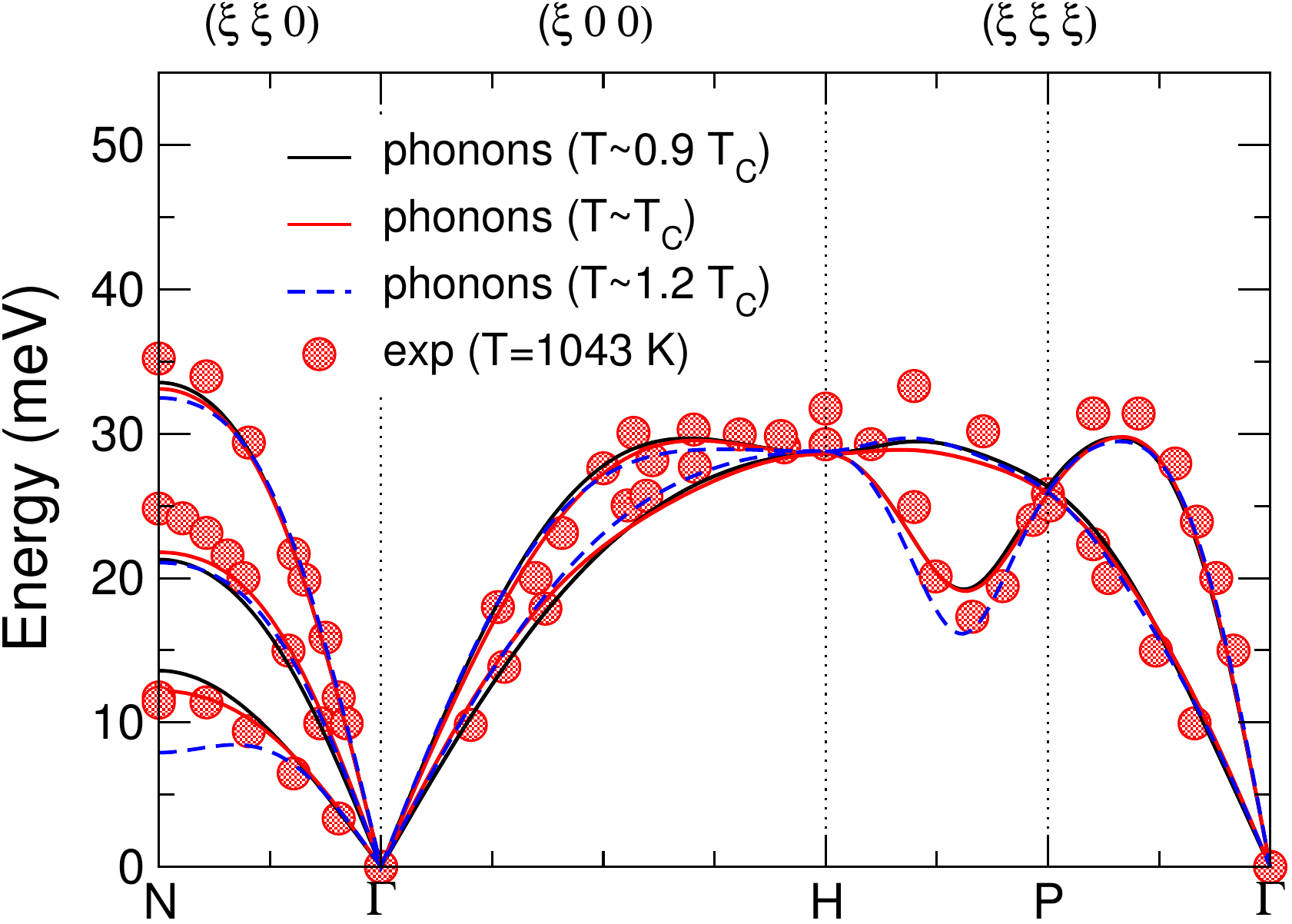}
      \includegraphics[width=0.5\textwidth,clip=true]{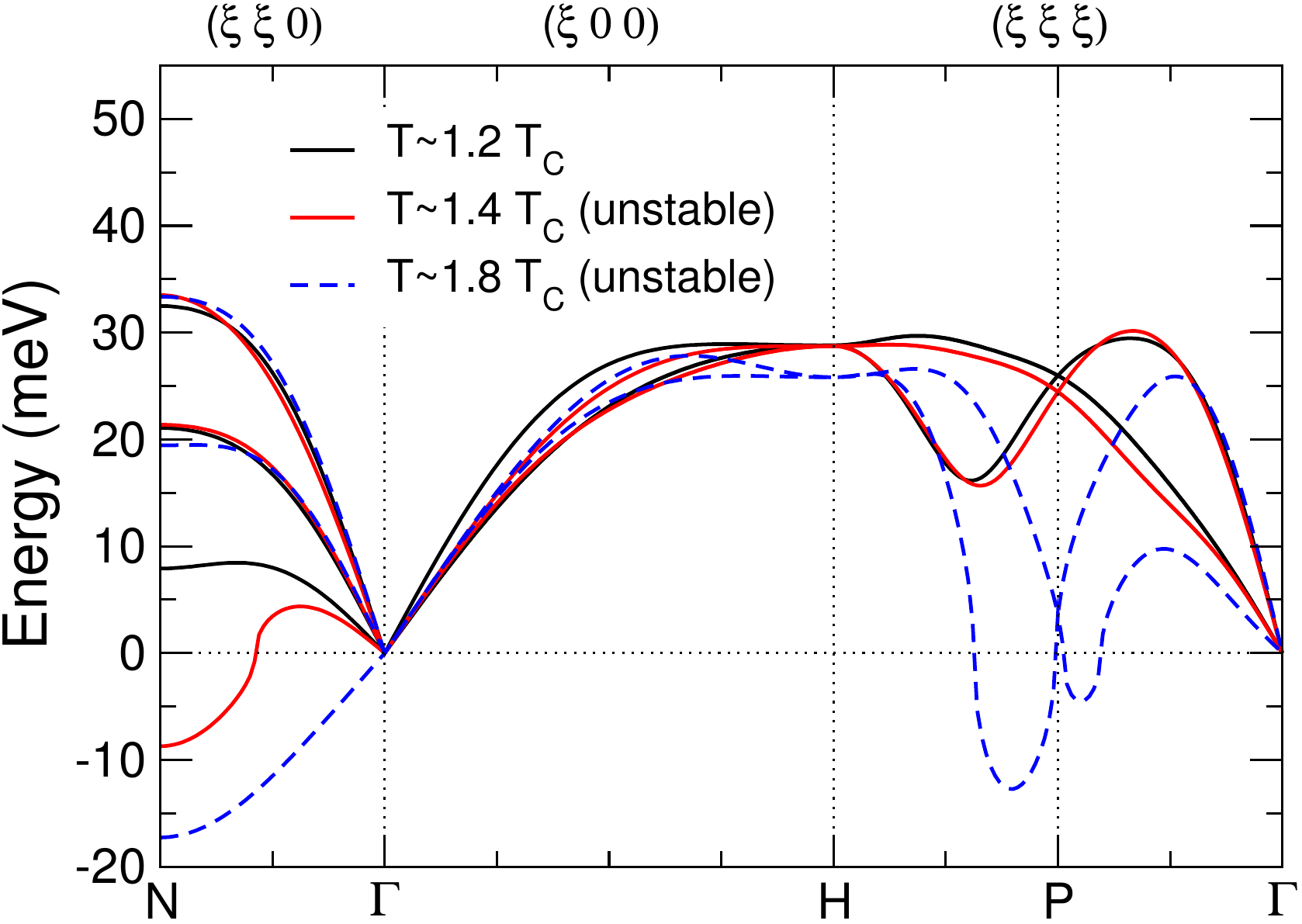}}
    \caption{Left panel: Calculated phonon dispersion curves for bcc iron near the Curie temperature. The results are compared with neutron inelastic scattering measurements at 1043~K~\cite{Neuhaus97}. Right panel: Calculated phonon dispersions of paramagnetic bcc iron near the $\alpha$-to-$\gamma$ and $\gamma$-to-$\delta$ phase transitions for different temperatures. From Ref.~\cite{Leonov14a} with permission of the authors.}
    \label{Fe:Fig_2}
  \end{figure}

  At even higher temperatures, above 1670 K, iron is known to make a structural transition
  to the $\delta$ phase, which is stable  up to the melting
  curve~\cite{Basinski55,Wyckoff63}. To explain the phase stability of $\delta$ iron, we calculate the
  temperature evolution of the phonon dispersions near the
  $\alpha$-to-$\gamma$ and $\gamma$-to-$\delta$ phase transformations
  within DFT+DMFT~\cite{Leonov14a}. The results for the
  phonon dispersion curves at temperatures near the magnetic phase transition
  temperature $T_C$ are compared in Fig.~\ref{Fe:Fig_2} (left). Phonon dispersion curves obtained in the
  temperature range of 1.4 -- 1.8 $T_C$ are also compared in
  Fig.~\ref{Fe:Fig_2} (right).

  The phonon frequencies are seen to be overall positive, both in the
  ferromagnetic and paramagnetic phase at around $T_C$. This implies
  the structural stability of the bcc phase in accordance with
  experiment. In the given temperature range the calculated phonon
  dispersions show only a rather weak temperature dependence. We
  notice a good quantitative agreement between the calculated phonon
  dispersions of bcc iron and the experimental data taken at $T \sim
  T_C$~\cite{Neuhaus97}.
  In particular, we find a remarkable anomaly in the
  transverse (110) acoustic mode ($T_1$ mode) along the $\Gamma$-$N$
  branch near the bcc-fcc phase transition temperature at $\sim$ 1.2
  $T_C$. As will be discussed below this behavior is a dynamical precursor of the
  bcc-to-fcc phase transition~\cite{Petry95} which occurs at $\sim$
  1.3 $T_C$.

  Upon further temperature increase our results, which were obtained in the
  harmonic approximation, clearly show that the bcc phase becomes
  dynamically unstable at $T \sim 1.4~T_C$, i.e., above the
  $\alpha$-to-$\gamma$ phase transition temperature~\cite{Leonov14a}. The origin of the
  instability lies in the $T_1$ mode in the $\Gamma$-$N$ direction discussed above,
  whose frequency becomes imaginary near the $N$-point. At the same
  time other phonon dispersion branches remain stable and are only weakly
  temperature dependent near the transition point. At even higher
  temperatures ($T \sim$ 1.8 $T_C$)  the bcc
  structure of the $\delta$ phase is dynamically unstable in the whole $\Gamma$-$N$
  branch, again due to the $T_1$ mode, with an additional anomaly near
  the $(\frac{2}{3},\frac{2}{3},\frac{2}{3})$ point. Therefore we
  expect that the $T_1$ transverse (110) phonon mode continuously
  softens upon temperature increase near the $\alpha$-$\gamma$ and
  $\gamma$-$\delta$ phase transition temperature -- a feature which is
  typical for simple nonmagnetic metals. By contrast, at temperatures
  $T < T_{\text{struct}}$ the phonon dispersion modes of bcc iron are
  relatively rigid, and resemble those of non-polymorphic bcc metals
  such as Cr, Mo, W. This should be contrasted with the phonon spectra of the denser fcc
  ($\gamma$) phase of iron which shows only a rather weak temperature
  dependence for all temperatures.

  An estimate of the lattice free energy,
  which is based on the quasi-harmonic method described in
  Ref.~\cite{Drummond02}, shows that the strong anharmonicity due to the
  $T_1$ mode at high temperatures is sufficient to overcome the total
  energy difference between the bcc and the fcc phases, resulting in the
  phase stability of the bcc $(\delta)$ phase~\cite{Leonov14a}. On the basis of these
  results we therefore conclude that the high-temperature bcc
  $(\delta)$ phase of iron is stabilized by the lattice entropy, which
  gradually increases upon heating due to the increasingly anharmonic
  behavior of the $T_1$ phonon mode.

\subsection{Correlation-induced topological Fermi surface transition in FeSe}
\label{sec:fese}

  As a second application of the DFT+DMFT scheme discussed in this section we address the electronic and structural properties of
  FeSe. This compound can be regarded as the parent compound of the
  Fe-based superconductors~\cite{kamihara08,Kamihara08_,Kamihara08__,Paglione10,Paglione10_,stewart11,Paglione10___}. It has been
  recently shown that the critical temperature $T_c$ of FeSe, which is
  about $T_c \sim$ 8 K at ambient pressure~\cite{Hsu08}, depends very
  sensitively on changes of the lattice structure of FeSe due to
  pressure or chemical doping. In particular, $T_c$ is found to
  increase up to $\sim 37$ K under hydrostatic pressure of about 7 GPa
  and to $\sim 14$ K upon chemical (isovalent) substitution with
  Te~\cite{Margandonna08,Margandonna08_,Mizuguchi08,Mizuguchi08_,Medvedev09,Sales09,Sales09_}.
  The angle-resolved photoemission and band structure calculations
  reveal that FeSe has the same Fermi surface topology as the
  pnictides~\cite{Xia09,Tamai10,Tamai10_,Tamai10__,Nakayama10,Nakayama10_,Nakayama10__,Nakayama10___,Nakayama10____,Nakayama10_____}. It is characterized by an
  in-plane magnetic nesting wave vector $Q_m = (\pi, \pi)$, consistent
  with $s_\pm$ pairing symmetry~\cite{Mazin08,chubukov08}. Moreover, both
  pnictides and chalcogenides display a strong enhancement of
  short-range spin fluctuations near $T_c$, with a resonance at $(\pi,
  \pi)$ in the spin excitation spectra~\cite{Christianson08,Christianson08_,Christianson08__,Christianson08___},
  suggesting a common origin of superconductivity in pnictides and
  chalcogenides, e.g., due to spin fluctuations~\cite{Mazin08,chubukov08}.
  On the other hand, the related isoelectronic compound FeTe exhibits
  no superconductivity~\cite{Medvedev09,Bao09,Bao09_,Bao09__}, but shows
  long-range antiferromagnetic  $(\pi,0)$ order below $T_N \sim$ 70 K,
  suggesting that the solid solution Fe(Se,Te) should have a remarkable
  crossover from the $(\pi, \pi)$ to $(\pi, 0)$ magnetic behavior upon
  substitution Se with Te. In addition, FeTe exhibits a remarkable
  phase transition under pressure, from a tetragonal to a
  collapsed-tetragonal phase, with a simultaneous collapse of local
  moments, indicating that the solid solution Fe(Se,Te) is close to an
  electronic and/or lattice transition~\cite{Li09,Li09_}.

  We will now discuss the origin of this surprising behavior as well as the
  properties of the Fe(Se,Te) solid solution, following the presentation by Leonov \emph{et al.}~\cite{Leonov15c}.
  Most recently these calculations were generalized to take into account full charge self-consistency~\cite{Skornyakov17}.
  It should be noted, however, that charge self-consistency is not essential here, i.e., qualitatively similar results are obtained for FeSe already within the non-charge self-consistent calculations reported earlier~\cite{Leonov15c}.

  The electronic structure and phase stability of FeSe is computed as a function
  of lattice volume, employing the DFT+DMFT approach
  implemented with plane-wave pseudopotentials~\cite{Leonov08,Leonov08_,Leonov08__}. For
  the partially filled Fe $3d$ and Se $4p$ orbitals we construct a
  basis set of atomic-centered symmetry-constrained Wannier
  functions~\cite{Anisimov05,Trimarchi08,Korotin08}. To solve the
  realistic many-body problem, we employ the continuous-time
  hybridization-expansion quantum Monte-Carlo
  algorithm~\cite{werner06,Gull11}. The calculations are performed at
  three different temperatures: $T =$ 290 K, 390 K, and 1160 K. In
  these calculations we use the average Coulomb interaction $\bar{U}
  =$ 3.5 eV and Hund's exchange $J =$ 0.85 eV, in accord with previous
  estimates for pnictides and
  chalcogenides~\cite{Haule08,Haule08_,Aichhorn09,Haule08___,Haule08____,Haule08_____,yin11,aichhorn11,Haule08________,Haule08_________,Haule08__________,Haule08___________,Haule08____________}. We employ the
  fully-localized double-counting correction, evaluated from the
  self-consistently determined local occupancies, to account for the
  electronic interactions already described by GGA. To investigate the
  phase stability, we take a tetragonal crystal structure (space group
  $P4/mmm$) with the lattice parameter ratio $c/a =$ 1.458 and Se
  position $z =$ 0.266, and calculate the total energy as a function
  of volume.

\subsubsection{Total energy and fluctuating moments}

\begin{figure}[tbp!]
    \centerline{
      \includegraphics[width=\textwidth,clip=true,viewport=0 550 590 784]{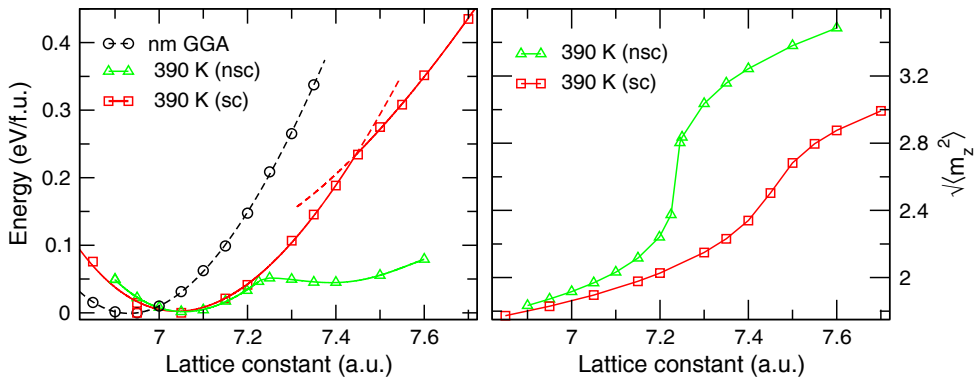}}
    \caption{%
    Total energy (left) and fluctuating local moment
     (right) of paramagnetic FeSe as a function of lattice
      constant calculated within DFT+DMFT at a temperature of 490 K
      with (sc) and without (nsc) charge self-consistency.
      The total energy is compared with that obtained within the non-magnetic GGA (``nm GGA'').
      After Ref.~\cite{Skornyakov17}.}
    \label{FeSe:Fig_1}
  \end{figure}

  Our results are presented in Fig.~\ref{FeSe:Fig_1}. In particular,
  we find the equilibrium lattice constant $a =$ 7.07 a.u., which is
  within 1\% of the experimental
  value~\cite{Margandonna08,Margandonna08_}.
  %
  %
  The calculated bulk modulus is $B \sim$
  70 GPa, which is comparable with that for iron pnictides. Most
  importantly,
  our DFT+DMFT calculations predict a structural
  transition of FeSe upon $\sim$ 10 \% expansion of the lattice
  volume. This result is unexpected and, in fact, is very different from that
  obtained with the non-spin-polarized GGA, where the bulk modulus comes out much higher.
  Once again the  results obtained within DFT+DMFT demonstrate the crucial importance
  of electronic correlations for the electronic structure and phase
  stability of FeSe. Namely,
the repulsive interaction leads to an increase of the unit cell volume
and hence results in a reduction of the bulk modulus.

  We interpret the structural transition  as a
  collapsed-tetragonal (low-volume) to tetragonal (high-volume) phase
  transformation upon expansion of the lattice volume. The phase
  transition is accompanied by a strong increase of the fluctuating
  local moment $\sqrt{\langle m^2_z\rangle}$, which grows
  monotonically upon expansion of the lattice. The high-volume phase
  has a much larger local moment and a softer lattice with a much
  lower bulk modulus of 35 GPa.

\subsubsection{Spectral function and Fermi surface structure}

  \begin{figure}[tbp!]
    \centerline{
      \includegraphics[width=0.95\textwidth,clip=true]{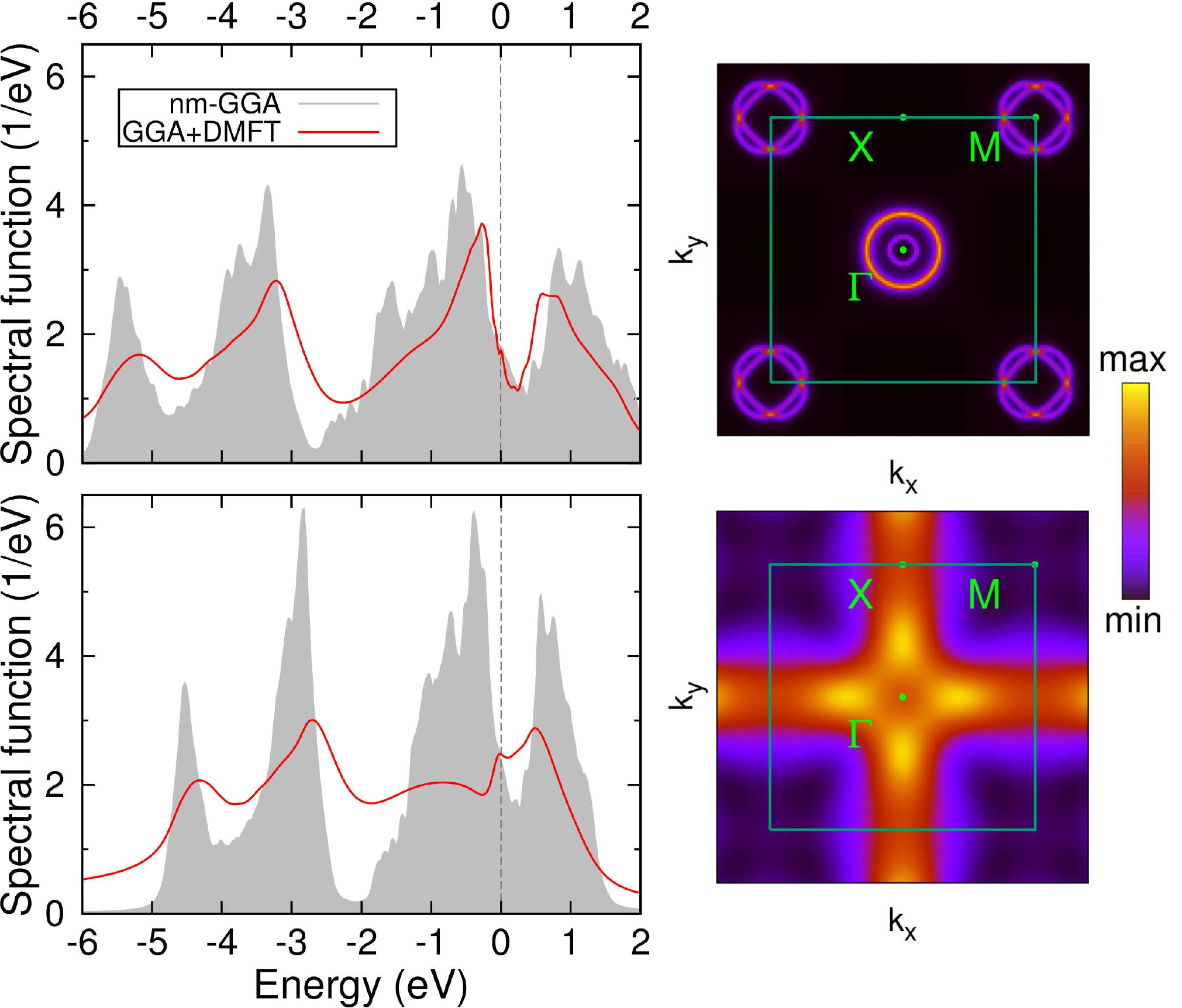}}
    \caption{Left column: Spectral functions of FeSe calculated
    using the  fully charge self-consistent DFT+DMFT scheme (lines)
    as compared with the results from the non-magnetic GGA (filled
      area). Right column: Fermi surface reconstruction in the (k$_x$, k$_y$) plane at k$_z$ = 0,
      calculated for paramagnetic FeSe
      using the fully charge self-consistent
      DFT+DMFT.
      Top row: Results obtained for $a=7.05$ a.u. (low volume). Bottom row:
      Results obtained for the expanded lattice (high volume), with $a=7.60$ a.u.
      After Ref.~\cite{Skornyakov17}.}
    \label{FeSe:Fig_2}
  \end{figure}

  The results for the integrated spectral function are shown in 
  Fig.~\ref{FeSe:Fig_2} (left). In agreement with previous
  studies~\cite{aichhorn11}, we notice a remarkable reduction of the
  Fe $3d$ bandwidth near the Fermi energy caused by electronic
  correlations. The lower Hubbard band is located at about $-1.5$ eV for
  both phases. Upon expansion of the lattice, we find a substantial
  spectral weight transfer. In particular, the spectral function for
  the low-volume phase exhibits a well-defined quasiparticle peak
  located below the Fermi level at $-0.19$ eV, which is absent in the
  high-volume phase. We note that the peak originates from the van
  Hove singularity of the Fe $xz/yz$ and $xy$ bands at the
  $M$-point. Moreover, we find a substantial qualitative change in the
  self-energy upon expansion of the lattice, resulting in a
  significant orbital-selective renormalization of the Fe $3d$
  bands (not shown here). The $xz/yz$ and $xy$ bands exhibit significantly stronger correlations than 
  the $z^2$ and $x^2-y^2$ bands.  While in the
  low-volume phase the quasiparticle mass enhancement is moderate,
  $\sim$ 2.0 -- 2.5, our calculations for the high-volume phase yield
  a strong renormalization $\sim$ 4 for the $xz/yz$ orbitals and
  $\sim$ 6 for the $xy$ orbitals. This shows in particular
  that the effect of orbital-selective correlations increases upon
  expansion of the lattice.

  Our results for the Fermi surface presented in
  Fig.~\ref{FeSe:Fig_2} (right) reveal a complete reconstruction of the
  electronic structure upon expansion of the lattice, resulting in a
  dramatic change of the Fermi surface topology (``Lifshitz
  transition'')~\cite{Leonov15c}. In particular, the Fermi surface at
  the $M$-point collapses, leading to a large square-like hole pocket
  around the $M$-point in the high-volume phase, in surprising analogy
  with the cuprates. In addition, the hole pockets around the
  $\Gamma$-point transform into incoherent spectral weight at the
  Fermi level along the $\Gamma$-$X$ direction. The reconstruction of
  the Fermi surface topology leads to a corresponding change of the
  magnetic correlations in FeSe. We find in-plane nesting with $Q_m =
  (\pi,\pi)$, connecting hole and electron parts of the Fermi surface,
  to be dominant in the low-volume phase. Upon expansion of the
  lattice, the Lifshitz transition sets in, resulting in the $(\pi,
  0)$-type magnetic correlations in the high-volume phase.

  Our findings suggest that the proximity of a van
  Hove singularity to the Fermi level strongly influences, or even induces,
  (unconventional)
  superconductivity in the chalcogenide
  FeSe$_{1-x}$Te$_x$ series~\cite{Leonov15c}.

\subsection{First-principles calculation of atomic forces and structural distortions}
\label{sec:forces}

So far the phase stability of strongly correlated materials was investigated by performing total-energy calculations within  DFT+DMFT.
 These calculations are very demanding even for
  simple materials, since they require the minimization of the total
  energy as a function of all atomic displacements. The computational effort therefore increases exponentially, which strongly
  limits the applicability of total-energy based techniques.
  This obstacle can be overcome by computing the complete set of
  interatomic forces using the Hellmann-Feynman theorem, whereby it becomes
  possible to compute the lattice structure even of complex materials.
  We conclude this section by formulating the DFT+DMFT approach for
  the calculation of interatomic forces and structural distortions in
  correlated materials based on the implementation of DFT+DMFT within
  the linear-response formalism~\cite{Leonov14b}; %
  see also a recent formulation based on a stationary 
  implementation of the DFT+DMFT functional~\cite{haule2016f,haule2016g}.
  The calculation of
  forces makes it possible to compute atomic displacements and
  equilibrium atomic positions and, hence, to explain the origin of
  lattice transformations caused by electronic correlations. Moreover,
  it allows one to determine the equilibrium lattice structure of
  correlated systems even in the vicinity of a Mott metal-insulator
  transition -- a computation which was not feasible up to now.

We discuss the DFT+DMFT scheme for the calculation of interatomic forces and
structural distortions in correlated materials following Ref.~\cite{Leonov14b}.
For  illustrative purposes we restrict our presentation to a discussion of
  structural transitions in a correlated model system, elemental
  (solid) hydrogen. By choosing different values of the local Coulomb
  interaction parameter $U$, we are able to explore the properties of solid
  hydrogen near a Mott metal-insulator phase transition.

The interatomic force acting on the atom $s$ is calculated as the
  first-order derivative of the total energy functional (\ref{eq:energy}):
  \begin{eqnarray}
    \label{eq:force_functional}
    F_s &=& F^s_\mathrm{DFT} - \delta_s \langle {\hat H_\mathrm{DFT}} \rangle
    + \sum_{m,\mymathbf{k}} \delta_s \epsilon^\mathrm{DFT}_{m,\mymathbf{k}}  \notag \\
    &&-\,\frac{1}{2} \sum_{imm',\sigma\sigma'}
    U^{\sigma \sigma'}_{mm'} \delta_s \langle \hat n_{im\sigma} \hat n_{im'\sigma'}\rangle - F^s_\mathrm{DC}.
  \end{eqnarray}
  Here $\delta_s \equiv d/d \bf{R}_s$ denotes the
  derivative with respect to the atomic position $\bf{R}_s$, and
  $F^s_\mathrm{DFT}$ is the force on the atom $s$ calculated
  within DFT.  Furthermore, $\delta_s \langle {\hat H_\mathrm{DFT}}
  \rangle$ is the thermal average of the force operator
  $\delta_s \hat H_{\text{DFT}}$, which leads to the Hellmann-Feynman
  contribution given by the first-order changes of the DFT Wannier
  Hamiltonian $\hat H_{\text{DFT}}$, plus the term arising from the explicit
  dependence of the local Green function on the atomic positions:
  \begin{eqnarray}
    \label{eq:kinetic_force}
    \delta_s \langle {\hat H_\mathrm{DFT}} \rangle &=& \langle { \delta_s \hat H_\mathrm{DFT}} \rangle
    +
    \Tr\sum_{\bm{k},i\omega_n} { \hat H^{\bm{k}}_\mathrm{DFT} \delta_s \hat G_{\bm{k}}(i\omega_n) e^{i\omega_n0+}}.
  \end{eqnarray}
  The derivative of the local Green function is given by
  \begin{eqnarray}
    \label{eq:gf_derivative}
    \delta_s \hat G_{\bm{k}}(\omega) =
    \hat G_{\bm{k}}(\omega) [ \delta_s \hat H^{\bm{k}}_\mathrm{DFT}
    + \delta_s \hat \Sigma(\omega) - \delta_s \mu ] \hat G_{\bm{k}}(\omega).
  \end{eqnarray}
  To compute the interatomic forces caused by the Coulomb interaction (the 4th term on the right-hand side of
  Eq.~\eqref{eq:force_functional}), we make use of
  the
  first-order derivative of the Galitskii-Migdal
  formula~\cite{Migdal,Galitskii58} for the interaction energy
  $ F^s_\mathrm{U} = -\frac{1}{2} \Tr \sum_{i\omega_n}
  [ \delta_s \hat \Sigma(i\omega_n) \hat G(i\omega_n) +
  \hat \Sigma(i\omega_n) \delta_s \hat G(i\omega_n)] e^{i\omega_n0+}$.
  Here we assume that the average Coulomb interaction $\bar U$ and
  Hund's rule coupling $J$ retain their values when the atomic positions
  change. The force operator $\delta_s \hat H_{\text{LDA}}$
  and the first-order change of the self-energy $\delta_s \hat \Sigma
  (\omega)$ are the two independent variables in the force functional
  [Eq.~\eqref{eq:force_functional}] which need to be evaluated to
  compute the interatomic forces~\cite{chem_pot}.

  To determine $\delta_s \hat H_{\text{DFT}}$, we generalize the projection
  scheme used to evaluate the DFT Wannier
  Hamiltonian~\cite{Anisimov05,Trimarchi08}. The former is based on
  the projection of the set of site-centered atomic-like
  trial-orbitals $|\phi_n \rangle$ on the Bloch functions $|\psi_{ik}
  \rangle$ of the selected bands with indices ranging from $N_a$ to
  $N_b$. In this way the force operator may be written as
  \begin{eqnarray}
    \label{eq:force_operator}
    (\delta_s \hat H^{\bm{k}}_\mathrm{DFT})_{nm} &=& \sum_{i=N_a}^{N_b} \langle \phi_n|\psi_{i \bf{k}} \rangle \langle \psi_{i \bf{k}}|\phi_m \rangle
    \;
    (\delta_s V_{i \bf{k}}^\mathrm{KS} + \delta_s V_{i \bf{k}}^\mathrm{Hxc}),
  \end{eqnarray}
  where $\delta_s V_{i \bf{k}}^\mathrm{KS}$ and $\delta_s V_{i
    \bf{k}}^\mathrm{Hxc}$ denote the first-order changes in the LDA
  Kohn-Sham and the Hartree and exchange-correlation potentials,
  respectively~\cite{basis}. Within the plane-wave
  pseudopotential approach~\cite{Baroni01,Giannozzi09} the Kohn-Sham contribution $\delta_s V_{i
    \bf{k}}^\mathrm{KS}$ can be calculated  as
  \begin{eqnarray}
    \label{eq:analitic_force_operator}
    \delta_s V_{i \bf{k}}^\mathrm{KS} &\propto& -i \sum_{\bm{G},\bm{G'}} c^*_{i,\bm{k}+\bm{G}}c_{i,\bm{k}+\bm{G'}}
    e^{-i(\bm{G}-\bm{G'}) \bm{R}_s}
    (\bm{G}-\bm{G'}) V_s^\mathrm{KS}(\bm{k}+\bm{G},\bm{k}+\bm{G'}),
  \end{eqnarray}
  where $V_s^\mathrm{KS}(\bm{G},\bm{G'})$ is the Kohn-Sham potential
  for atom $s$ (for details see Refs.~\cite{DalCorso93,DalCorso93_}). The
  contribution $\delta_s V_{i \bf{k}}^\mathrm{Hxc}$ is obtained from
  linear-response DFT calculations~\cite{DalCorso93,DalCorso93_}.

  To evaluate the change of the self-energy $\delta_s \hat
  \Sigma(\omega)$ we perform a functional derivative of the impurity
  Green function (we suppress the spin/orbital indices and assume
  a summation over repeated indices)
  \begin{eqnarray}
    \label{eq:variational_approach}
    \delta_s \hat G(\tau_1-\tau_2) = - \hat \chi(\tau_1,\tau_2,\tau_3,\tau_4)~
    \delta_s \hat {\mathcal{G}}^{-1}(\tau_3,\tau_4)
  \end{eqnarray}
  with
  \begin{eqnarray}
    \label{eq:susceptibility}
    \hat \chi(\tau_1,\tau_2,\tau_3,\tau_4) = \langle \mathcal{T}_{\tau} \hat c(\tau_1) \hat c^{\dagger}(\tau_2) \hat c^{\dagger}(\tau_3) \hat c(\tau_4)\rangle -
    \langle \mathcal{T}_{\tau}  \hat c(\tau_1) \hat c^{\dagger}(\tau_2) \rangle
    \langle \mathcal{T}_{\tau}  \hat c^{\dagger}(\tau_3) \hat c(\tau_4) \rangle,
  \end{eqnarray}
  and make use of the first-order derivative of the local Green function
  [Eq.~\eqref{eq:gf_derivative}].  Eqs.~\eqref{eq:gf_derivative}
  and \eqref{eq:variational_approach} are solved self-consistently by employing
  $\delta_s \hat {\mathcal G}^{-1} = \delta_s \hat G^{-1} + \delta_s
  \hat \Sigma$ and the two-particle correlation function, i.e., the
  generalized susceptibility, $\chi(\tau_1,\tau_2,\tau_3,\tau_4)$
  calculated within DMFT.
  As a starting point, we evaluate an initial guess for $\delta_s \hat
  \Sigma$ employing the static Hartree approximation
  \begin{eqnarray}
    \label{eq:sigma_guess}
    \delta_s \hat \Sigma \simeq U \delta_s \hat N &=& U \sum_{\bm{k},i\omega_n} \delta_s \hat G_{\bm{k}}(i \omega_n) e^{i \omega_n0+} \notag \\
    &=& U \sum_{\bm{k},i\omega_n} \hat G_{\bm{k}}(i \omega_n) [ \delta_s \hat H^{\bm{k}}_\mathrm{DFT} - \delta_s \hat \mu ] \hat G_{\bm{k}}(i \omega_n) e^{i \omega_n0+}.
  \end{eqnarray}
  This gives us the initial guess for $\delta_s \hat {\mathcal G}^{-1}$
  which we then insert into Eq.~\eqref{eq:variational_approach} to
  evaluate $\delta_s \hat G$. From $\delta_s \hat G$ and $\delta_s
  \hat {\mathcal G}$ we obtain a new estimate for $\delta_s \hat
  \Sigma$ which allows us to compute the new $\delta_s \hat G$ using
  the $\bm{k}$-integrated Dyson equation
  [Eq.~\eqref{eq:gf_derivative}]. This scheme is iterated until self-consistency over the $\delta_s \hat \Sigma$ is reached.
  %

\subsubsection{Applications to elemental solid hydrogen}

To test the formalism proposed here, we perform a series of
  calculations for the simplest correlated electron problem, namely
  elemental (solid) hydrogen assuming a cubic structure with lattice
  constant $a=8$ atomic units (a.u.).  We then compare our results for the
  total energy computed as a function of atomic displacement with
  those obtained by the numerical integration of the corresponding
  forces~\cite{Leonov14b}.
  Furthermore, by varying the local Coulomb interaction $U$, we
  explore the structural properties of solid hydrogen near a Mott
  metal-insulator phase transition. Thereby we can test whether our approach
  is able to determine structural transformations
  in the vicinity of the Mott-Hubbard transition, which is still a challenging
  problem of present-day solid state physics.

\begin{figure}[tbp!]
    \centerline{
      \includegraphics[width=0.75\textwidth,clip=true,viewport=0 405 532 764]{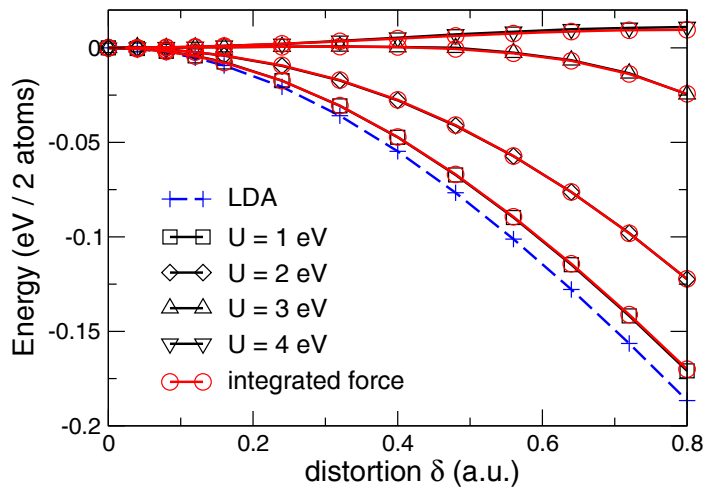}}
  \caption{%
    Comparison of the total energies of paramagnetic hydrogen
      computed by DFT+DMFT with the results obtained by (i) numerical
      integration of the corresponding force with respect to an atomic
      displacement, and (ii) within the LDA. The calculations are performed using different
      values of the Coulomb interaction $U$. Adapted from Ref.~\cite{Leonov14b} with permission of the authors.}
      \label{H:Fig_1}
  \end{figure}

  The nonmagnetic LDA calculations for cubic hydrogen yield a metallic
  solution with a half-filled hydrogen $s$ band of 3 eV width located at the
  Fermi level.  To evaluate the force, we consider a supercell with
  two hydrogen atoms, in which one of the atoms is displaced by a
  distance $\delta$ with respect to its crystallographic position. In
  Fig.~\ref{H:Fig_1} we show the results for the total energy
  obtained by LDA as a function of $\delta$. The nonmagnetic LDA
  calculations find the cubic lattice of hydrogen to be unstable since
  the total energy decreases with $\delta$.

  Now we take into account the electronic correlations by calculating the properties
  of paramagnetic hydrogen using the DFT+DMFT method. For the partially filled hydrogen $s$
  orbitals a basis of atomic-centered symmetry constrained Wannier functions is constructed.
 The calculations are performed for the $U$ values in the range of 1~--~4 eV
  at a temperature $T = 0.1$ eV.

  In Fig.~\ref{H:Fig_1} we also present our results for
  the total energy calculated by DFT+DMFT for paramagnetic hydrogen as
  a function of the displacement $\delta$. By changing the $U$ values,
  we were able to check the accuracy of our method by calculating the
  kinetic and interaction contributions, respectively, to the total
  force. By integrating the corresponding force with respect to
  $\delta$, we find excellent agreement (within 1~--~2
  meV) between the force-based  and the total energy calculations --- even for
  large displacements $\delta$ (up to $\sim$ 10 \% of the lattice
  constant $a$).
Most interestingly, by increasing $U$, the cubic lattice
  (more precisely, the investigated displacive mode) becomes (meta-)
  stable for $U$ $\geq$ $4$ eV.  These results clearly demonstrate the
  crucial importance of electronic correlations for the lattice
  stability of correlated materials.

Further applications of the DFT+DMFT scheme implemented within the
  linear-response formalism to SrVO$_3$ and
  KCuF$_3$ are discussed in Ref.~\cite{Leonov14b}. Our results for
  solid hydrogen, the correlated metal SrVO$_3$, and the correlated
  Mott-Hubbard insulator KCuF$_3$, demonstrate that the DFT+DMFT
  linear-response method presented here provides a robust
  computational tool for the study atomic displacements caused by
  electronic correlations. In particular, it allows one to determine
  the structural phase stability of both metallic and insulating
  correlated materials. The approach opens the way to calculate forces
  and thereby explore lattice instabilities induced by electronic
  correlations.
Lattice dynamical properties of correlated
  electron materials can also be calculated by implementing the approach
  with, for example, the so-called small displacements method (see,
  e.g., Ref.~\cite{Alfe09}).

\section{Conclusions and Outlook}
\label{sec:conclusions}

In this article we reviewed
the main results of our research into multi-band correlation phenomena
during the funding period of the DFG Research Unit FOR 1346.
In the first part of the paper we discussed the DMFT treatment of spontaneous symmetry breaking. We outlined the two general
approaches to this problem: first, monitoring the divergent susceptibilities in the normal phase that indicate an instability towards long range ordering
and, second, calculations in the phases with broken symmetry. As a pilot problem we studied materials close to the spin-state crossover. The investigation of the 
two-band Hubbard model, which provides a minimal description of this phenomenon, revealed a rich phase diagram with numerous phases
with diverse properties. The linear-response approach to the calculation of susceptibilities turned out to be very useful for the identification of
continuous phase transitions without any prior knowledge of the type of symmetry breaking.
Nevertheless calculations in the ordered phases are indispensable in order to identify first-order transitions or phase separation
as well as to investigate the properties of the ordered states.

Besides serving as a non-trivial test ground for the implemented formalism, these model calculations had a very specific materials based
motivation, namely the long-standing problem of spin-state crossover in compounds from the LaCoO$_3$ family. We performed a
series of DFT+DMFT studies of various aspects of the LaCoO$_3$ physics, the results of which led us to propose
a new picture of this material based on the concept of spinful excitons and their condensation. Experimental
efforts to test the proposal are under way.

The competition of spin states in the two-band Hubbard model and the resulting long-range order, including
excitonic magnetism and superconductivity, have been studied so far only on simple lattices, such as the Bethe and hypercubic lattice.
Other lattice geometries and hopping patterns will undoubtedly lead to new types of order,
e.g., the formation of a supersolid or magnetically ordered supersolid, which is of great theoretical interest.
Experimental realizations are ultimately necessary to test these results.
A first step is to establish experimental techniques capable of unambiguously identifying these exotic states of matter.
Scattering experiments, e.g., employing resonant inelastic x-ray scattering (RIXS), and their dependence on the polarization may
have this capability. Theoretical simulations of RIXS spectra in the different states are therefore necessary.

In the second part of the paper we reviewed the DFT+DMFT scheme for the computation of the electronic structure and phase stability of correlated materials.
This method allows one to explore structural transformations of the ionic lattice caused by correlations among the electrons, and to investigate lattice dynamical properties of correlated materials. We employed the
DFT+DMFT scheme to explain the phase
  stability and lattice dynamical properties of
elemental paramagnetic iron near the bcc-fcc phase
  transition. Furthermore, we discussed
  the origin of the  phase stability of the high-temperature $\delta$-phase. Our calculations clearly demonstrated the crucial
  importance of electronic correlations for the
  thermodynamic and lattice dynamical stability of the
  paramagnetic bcc phase of iron.

We also investigated the parent compound of the Fe-based superconductors, FeSe.   In particular, we computed the electronic
  structure and phase stability of FeSe as a function of lattice  volume and predicted a topological change (Lifshitz transition) of the Fermi surface upon expansion of the lattice as can be achieved experimentally by substituting Se by Te. This reconstruction is accompanied by a sharp increase of the local moments.

The present implementation of DFT+DMFT employs an adiabatic (Born-Oppenheimer) approximation, where it is assumed that the electrons move in an effective
potential produced by the heavy nuclei, i.e., the electrons and nuclei are treated independently. This simplification excludes a dynamic coupling
between the electronic and lattice degrees of freedom, which is of importance for the investigation of, for example, dynamical polaron effects.
In principle, appropriate generalizations of the DFT+DMFT scheme can be achieved by combining it with linear-response techniques to compute the  electron-phonon coupling and by the implementation of non-equilibrium DFT+DMFT methods employing the Keldysh formalism for the electrons
with (quantum) molecular dynamics simulation techniques~\cite{aoki14a}. We leave these projects for the future.

The field of DMFT has developed into many directions in the period covered here. Among them is the growing interest
in two-particle quantities which, after all, describe particularly important physical properties and also the technologically most exploited features of materials.
New impurity solvers were implemented~\cite{alps2,triqs,parragh12,lu14,wolf15}, and new algorithms developed~\cite{hafermann12,boehnke11,gunacker15,gunacker16} that greatly improved the computation of two-particle correlation functions. Along with the development of diagrammatic approaches
beyond DMFT~\cite{toschi07,rubtsov08,rubtsov12}, fundamental questions about two-particle consistency~\cite{vanloon16}, the consequences of its violation in DMFT, and their possible remedies were raised and are expected to give fresh impetus.
The availability of two-particle correlators for multi-orbital quantum impurities, which allows for the application of linear
response to realistic systems, e.g., with full $d$-shells, calls for further developments in the implementation of
this data-intensive formalism.

\section*{Acknowledgments}

We gratefully acknowledge support by the Deutsche Forschungsgemeinschaft through FOR 1346.

\bibliography{p2instabilities,p2lattice,p2footnotes,p2bibadd,p2bibmore,p2propbib}

\begin{thebibliography}{100}
\bibitem{correspondence}
This correspondence is only approximate for model with asymmetric band or with
  cross-hopping.
\bibitem{Migdal}
$F^s_\mathrm{U}$ can be also evaluated using a variation of the double
  occupancy matrix $\delta_s \langle \hat n_{im\sigma} \hat
  n_{im'\sigma'}\rangle$. However, this is computationally more demanding,
  since it requires the calculation of a three-particle correlation function.
\bibitem{basis}
We assume here that the Wannier basis is complete, i.e., there is no
  contribution caused by the change of the projectors $\delta_s \langle \phi_n
  | \psi_{i \bf{k}}\rangle $. In general, $\delta_s \langle \phi_n | \psi_{i
  \bf{k}}\rangle $ can be evaluated using the orthonormality conditions
  $\sum_{i=N_1}^{N_2}\langle \phi_n | \psi_{i \bf{k}}\rangle \langle \psi_{i
  \bf{k}} | \phi_m\rangle = \delta_{nm}$.
\end{thebibliography}
\bibliographystyle{spphys}

\end{document}